\documentclass[aps,prd,superscriptaddress,preprintnumbers,10pt]{revtex4-2}
 \usepackage{ifpdf}
\usepackage{graphicx}
\usepackage{amsmath}
\usepackage{amssymb}
\usepackage{color}
\usepackage{bm}
\usepackage{mathrsfs}
\usepackage{slashed}
\usepackage{slashed}

\begin{document}

\title{Photon emission from weakly magnetized neutral pions}
\date{21 May, 2026}
\author{Xinyang Wang}
\email{wangxy@aust.edu.cn}
\affiliation{Center for Fundamental Physics, School of Mechanics and Physics, Anhui University of Science and Technology, Huainan, Anhui 232001, China}

\author{Fan Lin}
\email{linfan@aust.edu.cn}
\affiliation{Center for Fundamental Physics, School of Mechanics and Physics, Anhui University of Science and Technology, Huainan, Anhui 232001, China}

\author{Igor Shovkovy}
\email{igor.shovkovy@asu.edu}
\affiliation{College of Integrative Sciences and Arts, Arizona State University, Mesa, Arizona 85212, USA}
\affiliation{Department of Physics, Arizona State University, Tempe, Arizona 85287, USA}

\begin{abstract}
Using a hadronic framework, we derive an explicit expression for photon production from neutral pions in a weak background magnetic field. Our calculation is built on the proton triangle diagram with an effective Yukawa $\pi^0$-proton coupling, offering an alternative to quark-level descriptions that is advantageous when the magnetic length greatly exceeds the proton size. Corrections to the pion decay constant are computed up to second order in the magnetic-field strength, revealing that the field generally suppresses the decay rate. Quantitatively, however, the effect remains modest even for fields as strong as $|eB|\simeq m_\pi^2$. The differential photon emission rate exhibits anisotropy, with the strongest suppression occurring when the pion momentum is perpendicular to the magnetic field. Overall, the modification of the $\pi^0 \to \gamma\gamma$ rate is parametrically small, scaling as $|eB|^2/m_P^4$, where $m_P$ is the proton mass. While the magnetic-field-induced anisotropy is conceptually interesting in principle, it is likely too small to be resolved in present heavy-ion measurements.
\end{abstract}
\maketitle

\section{Introduction}
\label{sec:introduction}

The decay of neutral pions into two photons ($\pi^0 \rightarrow \gamma \gamma$) is one of the most fundamental processes in quantum chromodynamics (QCD). It serves as a crucial test of chiral symmetry breaking and the axial anomaly within the Standard Model. From an experimental standpoint, this decay is commonly observed in high-energy physics and provides a primary signature for identifying neutral pions in both collider experiments and astrophysical observations, such as those involving cosmic rays. The latter, originating from supernova remnants, can produce pions through interactions with the interstellar medium. Among these, neutral pions, being unstable, decay within approximately $10^{-16}$ seconds into pairs of gamma-ray photons, each carrying roughly half of the pion's rest mass. The detection of such photons serves as a key observational signature, offering insights into the high-energy processes associated with supernova remnants \cite{Fermi-LAT:2013iui}. It is argued that the decay of neutral pions provides the dominant component of the truly diffuse $\gamma$-ray emission observed in the High Energy Stereoscopic System~\cite{HESS:2006vje}.

In the context of heavy-ion collisions, photons act as unique electromagnetic probes \cite{Rapp:2016xzw,Gale:2025ome}. Due to the long mean free path associated with electromagnetic interactions, photons traverse the medium with minimal rescattering, arriving at the detector largely undistorted. This property allows them to carry direct information about the spacetime evolution and dynamics of the medium at the moment of their production. Photons are emitted throughout all stages of a heavy-ion collision, including contributions from prompt photons, thermal radiation, jet-medium bremsstrahlung, and others \cite{Paquet:2015lta,Gale:2021emg}. In experiments, direct photons are defined as all photons not originating from hadron decays. Among hadronic decay sources, neutral pion decays account for approximately $80\%$ of the total decay photon yield \cite{PHENIX:2008uif}. Therefore, in data analysis, the total decay photon contribution is often estimated by scaling the measured pion decay photons by a factor of $10/8$.

In noncentral heavy-ion collisions, extremely strong magnetic fields are generated during the initial stages of the collision. These transient fields, primarily created by the motion of charged ions, can reach or exceed typical QCD energy scales $|eB|\gtrsim m_\pi^2$ \cite{Skokov:2009qp,Voronyuk:2011jd,Deng:2012pc,Bloczynski:2012en,Guo:2019mgh}. Theoretically, such fields are predicted to induce novel quantum phenomena, including the chiral magnetic and chiral separation effects \cite{Shovkovy:2021yyw,Li:2020dwr,Kharzeev:2024zzm}. More recently, studies have also shown that strong magnetic fields can influence the amplitudes of particle emission processes and contribute to anisotropies in electromagnetic observables \cite{Wang:2024gnh,Wang:2023fst,Wang:2022jxx,Wang:2020dsr}. For instance, the magnetic field can enhance the thermal production rates of electromagnetic probes, thereby affecting both the yield and angular distribution of direct photons \cite{PHENIX:2008uif,PHENIX:2014nkk,ALICE:2015xmh,ALICE:2017nce,PHENIX:2022rsx}. 

As the quark-gluon plasma expands and hadronizes, the magnetic field is expected to decay rapidly. Consequently, during the later stages of the collision, when pions become abundant, the field is likely to be substantially weaker. A quantitative estimate of its strength at that stage remains uncertain, however, because it depends sensitively on the poorly constrained electrical conductivity of the strongly interacting quark-gluon plasma and on possible departures from Ohm's law in a rapidly expanding medium. It is therefore important to quantify any magnetic-field-induced modification of neutral-pion decay, which may serve as an independent theoretical probe of the magnetic-field strength.

Although neutral pions themselves do not couple directly to magnetic fields, their decay processes can be influenced indirectly. In the presence of sufficiently strong magnetic fields, quantum electrodynamic effects, such as anomalous decay rate modifications, may become significant. These effects arise from field-induced changes in the properties of the vacuum, including modifications to the dispersion relations of photons and vacuum birefringence. As a result, the angular distribution and attenuation of the emitted photons can also be altered. In this work, we focus specifically on the anomalous modifications to the neutral pion decay rate induced by strong magnetic fields.

This paper is organized as follows. In Sec.~\ref{sec:formalism}, we present the theoretical formalism for calculating the decay of neutral pions into a pair of photons in the presence of a background magnetic field. In Sec.~\ref{sec:numerical-results}, we present our main numerical results. Finally, we summarize our main findings and discuss the limitations and possible implication of our results in Sec.~\ref{sec:summary}. Various technical details and derivations are presented in several appendixes at the end of the paper.

\section{Formalism}
\label{sec:formalism}

\begin{figure}[t]
\begin{center}
\includegraphics[width=0.5\textwidth]{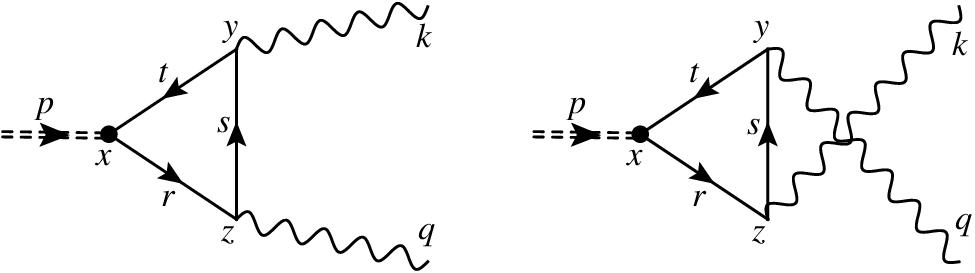}
\caption{Triangular one-loop Feynman diagrams contributing to the anomalous decay of the neutral pion into two photons.}
\label{pgg}
\end{center}
\end{figure}

The influence of a background magnetic field on pion decay, $\pi^0 \rightarrow \gamma \gamma$, may be analyzed within two complementary frameworks: a quark or a hadronic description. In both cases, the calculation necessarily involves the triangle diagrams shown in Fig.~\ref{pgg}, which embody the chiral anomaly. In the present work, we employ the hadronic description, following the standard textbook treatment in the absence of a magnetic field \cite{Schwartz:2014sze}. Although the quark-level formulation may offer certain advantages \cite{Weinberg:1996kr}, both approaches are expected to yield quantitatively similar results. Minor differences could arise from the distinct electric charges of the quarks and their direct coupling to the magnetic field. Our choice of the hadronic description is primarily motivated by our focus on the weak-field regime. In this case, over distance scales of the order of the magnetic length, $\ell = 1/\sqrt{|eB|}$, the proton can be accurately treated as a pointlike particle, and the details of its quark substructure are unlikely to play a significant role.

In the hadronic formulation, the fermion lines in the triangle diagrams of Fig.~\ref{pgg} correspond to protons, which carry electric charge. The interaction between the neutral pion and the proton is described by the Yukawa-type Lagrangian,
\begin{equation}
{\cal L}_{\rm int} = i \lambda  \pi^0 \left( \bar\psi_p \gamma^5 \psi_p\right),
\end{equation}
where the coupling constant is given by $\lambda = m_P/f_\pi$, with $m_P$ denoting the proton mass and $f_\pi \approx 92~\mbox{MeV}$ the vacuum pion decay constant.

In the present work, we employ a weak-field expansion organized in powers of the small parameter $|eB|/m_P^2$, where $m_P$ is the proton mass. Within this framework, magnetic-field-induced modifications of the proton mass and the vacuum pion decay constant are higher-order corrections.  Therefore, neglecting such effects is consistent with the accuracy of our calculation.

By definition, the decay rate is given by \cite{Schwartz:2014sze}
\begin{equation}
\label{decay-def}
\Gamma_{\mathbf{p}}(B) = \frac{1}{\sqrt{m_\pi^2+\mathbf{p}^2}}\int \frac{d^3 \mathbf{q}}{2 q_0 (2\pi)^3} \int \frac{d^3 \mathbf{k}}{2 k_0(2\pi)^3} 
\frac{|\mathcal{M}(k,q)|^2}{2 } ,
\end{equation}
where the amplitude for $\pi^0 \rightarrow \gamma\gamma$ decay is represented by the triangle diagrams in Fig.~\ref{pgg}. Note that $\mathbf{k}$ and $\mathbf{q}$ are the three-momenta of photons, and $\mathbf{p}=\mathbf{k}+\mathbf{q}$. The formal expression for the amplitude is given by
\begin{equation}
\label{first}
i\mathcal{M}(k,q)=-1(-\lambda)(-ie)^2 \epsilon_{\mu}^{1*}\epsilon_{\nu}^{2*} \sum_{i=1,2} M_i^{\mu\nu}(k,q),
\end{equation}
where $\epsilon_{\nu}^{i}$ are the photons polarization vectors and $M^{\mu\nu}_i(k,q)$ are the contributions from the two diagrams in Fig.~\ref{pgg}. By definition,
\begin{eqnarray}
\label{Mmunu1-1}
M^{\mu\nu}_1(k,q) &=&\int d^4 x d^4 y d^4 z \int \frac{d^4 r}{(2\pi)^4} \frac{d^4 s}{(2\pi)^4} \frac{d^4 t}{(2\pi)^4}e^{-ir\cdot (z-x) -is\cdot (y-z) -it\cdot (x-y)}e^{-ip \cdot x+i k\cdot y+i q \cdot z}\nonumber\\
&&\times \Phi(x,z)\Phi(z,y)\Phi(y,x) \text{Tr} \left [ \gamma^{\mu}\hat{G}(s)\gamma^{\nu}\hat{G}(r)\gamma^{5}\hat{G}(t)\right ],
\end{eqnarray}
and $M^{\mu\nu}_2(k,q) = M^{\nu\mu}_1(q, k)$. The explicit form of the proton propagator in a constant background magnetic field is given in Appendix~\ref{A1}. The factors $\Phi(x,z)$, $\Phi(z,y)$, and $\Phi(y,x)$ in Eq.~(\ref{Mmunu1-1}) are the Schwinger phases defined in Eq.~(\ref{Schwinger-phase}), which formally break the translation invariance of the charged fermion propagator. 

Without loss of generality, we choose the magnetic field $\mathbf{B}$ to point along the $+z$ direction and adopt the Landau gauge for the vector potential, $\mathbf{A} = (-By, 0, 0)$, where $B$ denotes the field strength. In this configuration, the product of the three Schwinger phases in Eq.~(\ref{Mmunu1-1}) takes the form
\begin{equation}
\Phi(x,z)\Phi(z,y)\Phi(y,x) = e^{\frac{i|e B|}{2}\epsilon^{ij}(y-x)_{i}(x-z)_{j}},
\label{prod-three-phases}
\end{equation}
where $e$ is the magnitude of the electron charge. Unlike the individual Schwinger phases, this product depends only on the relative spatial coordinates $y - x$ and $x - z$, and is therefore consistent with translation invariance in the plane perpendicular to the field.

In the weak-field regime, the translation-invariant part of Green's function can be expanded in powers of the magnetic-field strength, $eB$, as
\begin{equation}
\label{weak-iv}
\hat{G}(k_{\perp}, k_{\parallel}) = \hat{G}^{(0)}(k_{\perp}, k_{\parallel})+\hat{G}^{(1)}(k_{\perp}, k_{\parallel})+\hat{G}^{(2)}(k_{\perp}, k_{\parallel})+\cdots ,
\end{equation}
where the explicit forms of $\hat{G}^{(i)}(k_{\perp}, k_{\parallel})$ (with $i=0,1,2$) are provided in Appendix~\ref{A1}. Here, $k_{\perp}$ and $k_{\parallel}$ denote the momentum components transverse and parallel to the magnetic-field direction, respectively.
Generally, throughout this work, we employ the following notation for the metric tensors: 
$g^{\mu\nu} = \mathrm{diag}(1,-1,-1,-1)$, 
$g^{\mu\nu}_{\perp} = \mathrm{diag}(0,-1,-1,0)$, and 
$g^{\mu\nu}_{\parallel} = \mathrm{diag}(1,0,0,-1)$. 
Accordingly, we define the transverse and longitudinal components of the four-momentum as 
$k_{\perp} = (0, -k^1, -k^2, 0)$ and 
$k_{\parallel} = (k^0, 0, 0, -k^3)$, respectively. 
The corresponding invariants are 
$k_{\perp}^2 = -\mathbf{k}_{\perp}^2$, 
$\mathbf{k}_{\perp}^2 = k_x^2 + k_y^2$, 
$k_{\parallel}^2 = k_0^2 - k_3^2$, and 
$k^2 = k_{\parallel}^2 + k_{\perp}^2$.

The product of the Schwinger phases in Eq.~(\ref{prod-three-phases}) can likewise be expanded in powers of $eB$. Then, the contributions of the triangular diagrams to pion decay amplitude can be expressed via
\begin{equation}
\sum_{i=1,2} M_i^{\mu\nu}(k,q) \simeq  \mathcal{F}_{0}^{(0)}+\sum_{j=0}^{3}\mathcal{F}_{1}^{(j)}+\sum_{j=0}^{3}\mathcal{F}_{2}^{(j)} ,
\label{M_i}
\end{equation}
where the subscript denotes the order in $|eB|$, and the index $j$ distinguishes different contributions at the same order. The definitions and explicit forms of all terms are given in Appendix~\ref{App-Func}.
The resulting expressions for $\mathcal{F}_{\kappa}^{(j)}$, with $\kappa = 0, 2$, are
\begin{eqnarray}
\mathcal{F}_{0}^{(0)} &=&  4 m_P \epsilon^{\mu \nu \alpha \beta} k_{\alpha} q_{\beta} \mathcal{I}^{0}, \\
\mathcal{F}_{2}^{(0)} &=& -\frac{m_P  |e B|^2}{4(2\pi)^2} \epsilon^{\mu\nu\alpha\beta}\left[k_{\parallel,\alpha} q_{\parallel,\beta}\mathcal{K}_0 +k_{\parallel,\alpha} \delta_{1,\beta}\mathcal{K}_{q_x}+k_{\parallel,\alpha} \delta_{2,\beta}\mathcal{K}_{q_y} +q_{\parallel,\beta} \delta_{1,\alpha}\mathcal{K}_{k_x} \right.\nonumber\\
&+& q_{\parallel,\beta} \delta_{2,\alpha}\mathcal{K}_{k_y}+\delta_{2,\alpha}\delta_{1,\beta}\mathcal{K}_{q_x k_y}+\left.\delta_{1,\alpha}\delta_{2,\beta}\mathcal{K}_{q_y k_x}\right], \\
\sum_{j=1}^{3}\mathcal{F}_{2}^{(j)}  &=& 16 |e B|^2  m_P \epsilon^{\mu\nu\alpha\beta} k_{\alpha} q_{\beta} \sum_{i=1}^{3} \left(\mathcal{J}_i^1 + \mathcal{J}_i^2\right).
\end{eqnarray}
whereas all first-order terms vanish, $\mathcal{F}_{1}^{(j)} = 0$. The absence of terms linear in the magnetic field at the level of the decay amplitude may appear surprising at first sight. In fact, it follows from a generalized Furry-theorem argument, since such terms correspond to a closed proton loop with an odd number of electromagnetic vector insertions.

Indeed, a first-order correction in the background field amounts to one additional electromagnetic insertion on the triangle loop. The corresponding loop then contains three vector couplings in total: two associated with the outgoing photons and one with the background field, in addition to the $C$-even pseudoscalar pion vertex. Under reversal of the fermion flow, each vector vertex changes sign, whereas the pseudoscalar vertex remains unchanged. Because the number of vector insertions is odd, the loop amplitude changes sign, and the contributions from the two opposite fermion-flow orientations cancel exactly. In other words, this means that the effective action in the $C$-even pion sector contains only terms with an even total number of electromagnetic fields. Therefore, the $\pi^0\gamma\gamma A_{\mu,{\rm ext}}$ vertex is absent, whereas the $\pi^0\gamma\gamma A_{\mu,{\rm ext}}A_{\nu,{\rm ext}}$ vertex is allowed. This provides the microscopic explanation for why the first nonvanishing correction to the two-photon decay amplitude is quadratic in $B$. The explicit cancellations presented in Appendix~\ref{App-Func} confirms this general argument.

By making use of the expansion in powers of $eB$, we write the amplitude squared as follows:
\begin{eqnarray}
\label{am-F}
\frac{|\mathcal{M}(k,q)|^2}{2} &=& 4 \lambda^2 e^4 m_\pi^4 m_P^2 \left[\mathcal{I}^{0}+4|eB|^2\sum_{i=1}^{3} \left(\mathcal{J}_i^1 + \mathcal{J}_i^2\right) \right]^2  +\frac{\lambda^2 m_P^2 e^4|eB|^4}{4(2\pi)^4}\bigg\{\left(\mathcal{K}_0\right)^2\left[k_\parallel^2 q_\parallel^2 -(q_\parallel \cdot k_\parallel)^2 \right] \nonumber\\
&+& \left[(\mathcal{K}_{q_x})^2+(\mathcal{K}_{q_y})^2\right]k_\parallel^2+\left[(\mathcal{K}_{k_x})^2+(\mathcal{K}_{k_y})^2\right]p_\parallel^2+(\mathcal{K}_{q_x k_y}-\mathcal{K}_{q_y k_x})^2 - 2(\mathcal{K}_{q_x}\mathcal{K}_{k_x}+\mathcal{K}_{q_y}\mathcal{K}_{k_y})(k_\parallel \cdot q_\parallel)\bigg\},\nonumber\\
&&
\end{eqnarray}
where all $\mathcal{K}_{s}$ functions are defined in Appendix~\ref{App-Func}. Given the cumbersome analytic structure of these functions, it is convenient to simplify the analysis by considering several physically relevant approximations:
(i) the center-of-mass (CM) frame, defined by $\mathbf{k} = -\mathbf{q}$;
(ii) the small-pion-mass (SP) approximation, corresponding to $m_\pi \ll m_P$; and
(iii) the combined case (CS), which incorporates both the center-of-mass frame and the small-pion-mass approximation.

\subsection{Center-of-mass frame with the small-pion-mass approximation (CS)}

The main result for the amplitude in Eq.~(\ref{am-F}), obtained within a weak-field expansion in the small parameter $eB/m_P^2$, can be evaluated numerically without further approximation. For additional analytical insight, it is convenient to exploit the hierarchy $m_\pi \ll m_P$. Treating the pion mass as a second small parameter, we derive a simplified analytical expression for the decay rate.

Let us start from the simplest case of the center-of-mass frame, which is obtained by setting $\mathbf{k} = -\mathbf{q}$, and combine it with the small-pion-mass approximation,  $m_\pi\ll m_P$. In this case, from Eq.~(\ref{M_i}), we obtain 
\begin{eqnarray}
\sum_{i=1,2} M_i^{\mu\nu,\text{CS}}(k,q) &\hspace{-12pt}&=\frac{4}{m_P}\epsilon^{\mu \nu \alpha \beta} k_{\alpha} q_{\beta} \left(\frac{-i}{16\pi^2}\right)+ i|e B|^2  m_P \epsilon^{\mu\nu\alpha\beta} k_{\alpha} q_{\beta}\left[ \frac{1}{20\pi^2 m_P^6}+\frac{ \mathbf{q}_\perp^2}{126\pi^2 m_P^8} \right]-\frac{m_P  |e B|^2}{2(2\pi)^2} \epsilon^{\mu\nu\alpha\beta}\Bigg[k_{\parallel,\alpha} q_{\parallel,\beta} \nonumber\\
&\hspace{-12pt}&\times\frac{8 \mathbf{q}_\perp^2 -14 m_P^2}{105 m_P^{8}}+k_{\parallel,\alpha} q_{\perp,\beta}\frac{8\mathbf{q}_\perp^2-28m_P^2}{105 m_P^{8}}+ k_{\perp,\alpha} q_{\parallel,\beta}\frac{8\mathbf{q}_\perp^2-28m_P^2}{105 m_P^{8}}k_{\perp,\alpha}q_{\perp,\beta}\frac{140m_P^2-24\mathbf{q}_\perp^2}{315m_P^8}\Bigg],
\end{eqnarray}
where we used the results in Eqs.~(\ref{Final-Vaccum}), (\ref{Final-F20}), and (\ref{Final-F02}) from Appendix~\ref{App-Func}. Furthermore, by using the definition in Eq.~(\ref{first}), the corresponding amplitude becomes
\begin{eqnarray}
\mathcal{M}^{\text{CS}}&=&i\lambda e^2 \epsilon_{\mu}^{1*}\epsilon_{\nu}^{2*} \sum_{i=1,2} M_i^{\mu\nu,\text{CS}}(k,q)\nonumber\\
&=&\lambda e^2 \epsilon_{\mu}^{1*}\epsilon_{\nu}^{2*}\epsilon^{\mu \nu \alpha \beta} \left\{\frac{1}{4\pi^2 m_P} k_{\alpha} q_{\beta}\left(1-\frac{|eB|^2}{5m_P^4}-\frac{2|eB|^2 \mathbf{q}_\perp^2}{63m_P^6}\right) -\frac{i m_P  |e B|^2}{2(2\pi)^2}\left  [k_{\parallel,\alpha} q_{\parallel,\beta} \frac{8 \mathbf{q}_\perp^2 -14 m_P^2}{105 m_P^{8}}\right.\right.\nonumber\\
&+&\left. \left.  k_{\parallel,\alpha}  q_{\perp,\beta}\frac{4(2\mathbf{q}_\perp^2-7m_P^2)}{105 m_P^{8}}+k_{\perp,\alpha} q_{\parallel,\beta} \frac{4(2\mathbf{q}_\perp^2-7m_P^2)}{105 m_P^{8}}+k_{\perp,\alpha}q_{\perp,\beta}\frac{4(35m_P^2-6\mathbf{q}_\perp^2)}{315m_P^8}\right]\right\},
\end{eqnarray}
and its complex conjugate expression reads
\begin{eqnarray}
\mathcal{M}^{*,\text{CS}}&=&\lambda e^2 \epsilon_{\sigma}^{1}\epsilon_{\rho}^{2}\epsilon^{\sigma \rho \gamma \theta} \left\{\frac{1}{4\pi^2 m_P} k_{\gamma} q_{\theta}\left(1-\frac{|eB|^2}{5m_P^4}-\frac{2|eB|^2 \mathbf{q}_\perp^2}{63m_P^6}\right) +\frac{i m_P |e B|^2}{2(2\pi)^2}\left[k_{\parallel,\gamma} q_{\parallel,\theta}\frac{8 \mathbf{q}_\perp^2 -14 m_P^2}{105 m_P^{8}}\right.\right.\nonumber\\
&+&\left. \left. k_{\parallel,\theta} q_{\perp,\gamma}\frac{4(2\mathbf{q}_\perp^2-7m_P^2)}{105 m_P^{8}}+k_{\perp,\theta}q_{\parallel,\gamma} \frac{4(2\mathbf{q}_\perp^2-7m_P^2)}{105 m_P^{8}}+k_{\perp,\theta}q_{\perp,\gamma}\frac{4(35m_P^2-6\mathbf{q}_\perp^2)}{315m_P^8}\right]\right\}.
\end{eqnarray}
Using the on-shell conditions $k^2 = q^2 = 0$ and $q \cdot k = m_\pi^2/2$, and noting that 
\begin{eqnarray}
&&\epsilon_{\mu}^{1*}\epsilon_{\nu}^{2*}\epsilon^{\mu \nu \alpha \beta} \left[A_1 k_{\alpha} q_{\beta}+ i \left(B_1  k_{\parallel,\alpha}q_{\parallel,\beta} +B_2  k_{\parallel,\alpha} q_{\perp,\beta} + B_3  k_{\perp,\alpha} q_{\parallel,\beta}+B_4 k_{\perp,\alpha}q_{\perp,\beta}\right)\right]\nonumber\\
&& \times\epsilon_{\sigma}^{1*}\epsilon_{\rho}^{2*}\epsilon^{\sigma \rho \gamma \theta} \left[A_1 k_{\gamma} q_{\theta}- i \left(B_1  k_{\parallel,\gamma}q_{\parallel,\theta} +B_2  k_{\parallel,\gamma} q_{\perp,\theta} + B_3  k_{\perp,\gamma} q_{\parallel,\theta}+B_4 k_{\perp,\gamma}q_{\perp,\theta}\right)\right]
=2\Big\{\frac{m_\pi^4}{4}A_1^2
\nonumber\\
&&-B_1^2[k_\parallel^2 q_\parallel^2-(q_\parallel\cdot k_\parallel)^2]-B_2^2(k_{\parallel}^2q_\perp^2)-B_3^2(k_\perp^2q_{\parallel}^2)-B_4^2[k_\perp^2 q_\perp^2-(q_\perp\cdot k_\perp)^2]  +   2 B_2 B_3[(k_\perp \cdot q_\perp)(q_\parallel \cdot k_\parallel)]\Big\},
\end{eqnarray}
we obtain the following expression for the squared amplitude:
\begin{eqnarray}
\label{am-cm}
\frac{|\mathcal{M}^{\text{CS}}|^2}{2} &=&  \frac{\alpha_e^2 \lambda^2 m_\pi^4}{4\pi^2 m_P^2}\left(1-\frac{|eB|^2}{5m_P^4}-\frac{2|eB|^2 \mathbf{q}_\perp^2}{63m_P^6}\right)^2+\frac{\alpha_e^2 \lambda^2 m_P^2 |eB|^4}{4 \pi^2}\Bigg\{m_\pi^2 q_3^2 \left(\frac{8 \mathbf{q}_\perp^2 -14 m_P^2}{105 m_P^{8}}\right)^2 \nonumber\\
&& + 4 \mathbf{q}_\perp^2(\mathbf{q}_\perp^2+q_3^2)\left(\frac{4(2\mathbf{q}_\perp^2-7m_P^2)}{105 m_P^{8}}\right)^2\Bigg\},
\end{eqnarray}
where $\alpha_e$ is the fine structure constant.
In deriving this result, we used $\mathbf{k} = -\mathbf{q}$ and  $k^{0} = q^{0} = m_\pi/2$. Also, the following relations were employed:
\begin{eqnarray}
k_\parallel^2 q_\parallel^2-(q_\parallel\cdot k_\parallel)^2 &=& -(k^3 q_0 - k_0 q^3)^2 = -m_\pi^2 q_3^2,\\
k_\perp^2 q_\perp^2 -(k_\perp\cdot q_\perp)^2 &=& (k_y q_x -k_x q_y)^2 = 0, \\
B_2^2(k_{\parallel}^2q_\perp^2)+B_3^2(k_\perp^2q_{\parallel}^2)-2 B_2 B_3[(k_\perp \cdot q_\perp)(q_\parallel \cdot k_\parallel)] &=& -\bm{q}_\perp^2 \left[\bm{q}_\perp^2(B_2+B_3)^2+ 4 B_2 B_3 q_3^2\right].
\end{eqnarray}
Finally, by substituting Eq.~(\ref{am-cm}) into Eq.~(\ref{decay-def}), we obtain 
\begin{equation}
\Gamma^{\text{CS}}(B) \approx \Gamma(0)\left(1- \frac{2|eB|^2}{5m_P^4}-\frac{2 |eB|^2 m_\pi^2}{189m_P^6}\right),
\label{Gamma-CS}
\end{equation}
where $\Gamma(0)$ is the decay rate for the zero-field case
\begin{equation}
\Gamma(0) = \frac{\alpha_e^2 \lambda^2 m_\pi^3}{64 \pi^3 m_P^2} = \frac{\alpha_e^2 m_\pi^3}{64 \pi^3 f_\pi^2}, 
\end{equation}
which agrees with the textbook result in Ref.~\cite{Schwartz:2014sze}.  

It should be emphasized that the third term in the parentheses in Eq.~(\ref{Gamma-CS}) is not an independent higher-order correction. As is clear from the derivation, the auxiliary expansion in the small pion mass neither replaces nor invalidates the weak-field expansion. Instead, it provides a numerical approximation to the ${\cal O}(|eB|^2)$ result by exploiting the hierarchy $m_\pi\ll m_P$. Strictly speaking, this procedure is  self-consistent when $|eB|$ remains small compared to both $m_P^2$ and $m_\pi^2$. In our estimates below, we extrapolate the result beyond this strict regime of validity. However, its order-of-magnitude reliability is supported by the full numerical calculation, which does not rely on the small-pion-mass approximation.

As seen from Eq.~(\ref{Gamma-CS}), the qualitative effect of the magnetic field is to suppress the pion decay rate. Quantitatively, however, the effect remains very small even for relatively strong benchmark fields. For example, for magnetic fields of order $|eB| \simeq m_\pi^2\approx 0.02~\text{GeV}^2$, the correction to the decay rate is only about $0.02\%$. This value should be regarded as an optimistic benchmark rather than a realistic estimate for the stages of a heavy-ion collision in which neutral pions are abundant, since the magnetic field at such late times is expected to be smaller and the effect correspondingly more suppressed. Even for substantially stronger fields, $|eB| \simeq 10 m_\pi^2\approx 0.2~\text{GeV}^2$, the modification of the decay rate reaches only approximately $2\%$.

The reduction of the decay rate in the presence of a magnetic background is qualitatively consistent with results obtained using alternative approaches, such as those based on the Gell-Mann-Oakes-Renner relation~\cite{Shushpanov:1997sf,Agasian:2001ym,Simonov:2015xta,Coppola:2025nus} and chiral perturbation theory~\cite{Andersen:2012zc,Andersen:2012dz,Brauner:2017uiu}. These studies typically focus on the magnetic-field dependence of the pion decay constant, $f_{\pi}(B)$, which increases with the field strength. Since the decay rate scales as $\Gamma(B)\propto 1/f_{\pi}^2(B)$, an enhancement of $f_{\pi}(B)$ naturally leads to a suppression of $\Gamma(B)$. 

Nevertheless, our quantitative results differ from those reported in Refs.~\cite{Shushpanov:1997sf,Agasian:2001ym,Simonov:2015xta,Andersen:2012zc,Andersen:2012dz,Brauner:2017uiu}. As a useful benchmark, it is instructive to compare our findings with those of Ref.~\cite{Coppola:2025nus}, which employs a conceptually similar microscopic approach formulated at the quark level. We find that our magnetic-field-induced correction is approximately an order of magnitude smaller. For example, at $eB \simeq 0.1~\text{GeV}^2$, the authors of Ref.~\cite{Coppola:2025nus} find a reduction of the decay rate by about $15\%$  relative to its vacuum value, whereas our result yields only about a $1\%$ suppression. Most likely, this discrepancy reflects the very different dynamical frameworks underlying the two calculations, in particular our use of a hadronic description that does not resolve the internal structure of the proton. Clarifying the origin of these sizable quantitative differences will require a more detailed comparative analysis and should be addressed in future work.
  
\subsection{Small $m_\pi$ limit and arbitrary pion momentum (SP)}

Let us now consider the small $m_\pi$ approximation but lift the restriction of the center-of-mass frame, allowing the neutral pion to have an arbitrary momentum. In this case, by using Eqs. (\ref{Final-Vaccum}), (\ref{F20A-F}) and (\ref{K_smpi}), we have 
\begin{eqnarray}
\sum_{i=1,2} M_i^{\mu\nu, \text{SP}}(k,q) \approx\frac{4}{m_P}\epsilon^{\mu \nu \alpha \beta} k_{\alpha} q_{\beta} \left(\frac{-i}{16\pi^2}\right)+ i|e B|^2  m_P \epsilon^{\mu\nu\alpha\beta} k_{\alpha} q_{\beta}\left[ \frac{1}{20\pi^2 m_P^6}+\frac{ \mathbf{q}_\perp^2+\mathbf{q}_\perp\cdot\mathbf{k}_\perp+\mathbf{k}_\perp^2}{126\pi^2 m_P^8} \right]-\frac{m_P  |e B|^2}{2(2\pi)^2} &&\nonumber\\
\times \epsilon^{\mu\nu\alpha\beta}\left\{k_{\parallel,\alpha} q_{\parallel,\beta}\mathcal{K}_0^{\text{SP}}+k_{\parallel,\alpha} \delta_{1,\beta}\mathcal{K}_{q_x}^{\text{SP}}+k_{\parallel,\alpha} \delta_{2,\beta} \mathcal{K}_{q_y}^{\text{SP}}+q_{\parallel,\beta} \delta_{1,\alpha} \mathcal{K}_{k_x}^{\text{SP}}+q_{\parallel,\beta} \delta_{2,\alpha} \mathcal{K}_{k_y}^{\text{SP}}+\delta_{2,\alpha}\delta_{1,\beta} \mathcal{K}_{q_x k_y}^{\text{SP}}+\delta_{1,\alpha}\delta_{2,\beta}\mathcal{K}_{q_y k_x}^{\text{SP}}\right\}. &&\nonumber\\
\end{eqnarray}
By noting that 
\begin{eqnarray}
\epsilon_{\mu}^{1*}\epsilon_{\nu}^{2*}\epsilon^{\mu \nu \alpha \beta} \left[A_1 k_{\alpha} q_{\beta}+ i \left(B_1  k_{\parallel,\alpha}q_{\parallel,\beta} +B_2  k_{\parallel,\alpha} \delta_{1,\beta} + B_3  k_{\parallel,\alpha} \delta_{2,\beta}+B_4  q_{\parallel,\beta} \delta_{1,\alpha} + B_5  q_{\parallel,\beta} \delta_{2,\alpha}+B_6  \delta_{2,\alpha} \delta_{1,\beta} + B_7  \delta_{1,\alpha} \delta_{2,\beta}\right)\right]&& \nonumber\\
\times\epsilon_{\sigma}^{1*}\epsilon_{\rho}^{2*}\epsilon^{\sigma \rho \gamma \theta} \left[A_1 k_{\gamma} q_{\theta}- i \left(B_1  k_{\parallel,\gamma}q_{\parallel,\theta} +B_2  k_{\parallel,\gamma} \delta_{1,\theta} + B_3  k_{\parallel,\gamma} \delta_{2,\theta}+B_4  q_{\parallel,\theta} \delta_{1,\gamma} + B_5  q_{\parallel,\theta} \delta_{2,\gamma}+B_6  \delta_{2,\gamma} \delta_{1,\theta} + B_7  \delta_{1,\gamma} \delta_{2,\theta}\right)\right]&&\nonumber\\
=2\left\{ \frac{A_1^2 m_\pi^2}{4}-B_1^2[k_\parallel^2 q_\parallel^2-(q_\parallel\cdot k_\parallel)^2]-(B_2^2+B_3^2)k_{\parallel}^2-(B_4^2+B_5^2)q_{\parallel}^2-(B_6-B_7)^2+ 2 (B_2 B_4+ B_3 B_5)(k_\parallel \cdot q_\parallel)\right\},&&
\end{eqnarray}
we obtain the following result for the squared amplitude:
\begin{eqnarray}
\label{am-F-24}
\frac{|\mathcal{M}^{\text{SP}}|^2}{2} &\hspace{-12pt}&= \frac{\lambda^2 e^4 m_\pi^4}{4}\left[\frac{1}{4\pi^2 m_P}\left(1-\frac{|eB|^2}{5m_P^4}-\frac{2|eB|^2 (\mathbf{q}_\perp^2+\mathbf{q}_\perp\cdot\mathbf{k}_\perp+\mathbf{k}_{\perp}^2)}{63m_P^6}\right)\right]^2+\frac{\lambda^2 m_P^2 e^4|eB|^4}{4(2\pi)^4}\left \{\left(\mathcal{K}_0^{\text{SP}}\right)^2\left[k_\parallel^2 q_\parallel^2 -(q_\parallel \cdot k_\parallel)^2 \right]\right. \nonumber\\
&\hspace{-12pt}&+\left.\left[(\mathcal{K}_{q_x}^{\text{SP}})^2+(\mathcal{K}_{q_y}^{\text{SP}})^2\right]k_\parallel^2+\left[(\mathcal{K}_{k_x}^{\text{SP}})^2+(\mathcal{K}_{k_y}^{\text{SP}})^2\right]p_\parallel^2+(\mathcal{K}_{q_x k_y}^{\text{SP}}-\mathcal{K}_{q_y k_x}^{\text{SP}})^2 - 2(\mathcal{K}_{q_x}^{\text{SP}}\mathcal{K}_{k_x}^{\text{SP}}+\mathcal{K}_{q_y}^{\text{SP}}\mathcal{K}_{k_y}^{\text{SP}})(k_\parallel \cdot q_\parallel)\right\}.
\end{eqnarray}
The corresponding expression for the decay rate is then given by
\begin{equation}
\Gamma_{\mathbf{p}}^{\text{SP}} (B) \approx \Gamma_{\mathbf{p}}^{\text{SP}}(0)\left(1- \frac{2|eB|^2}{5m_P^4}- \frac{64 m_{\pi}|eB|^2}{63 \pi^2 m_P^6}\mathcal{A}_{\mathbf{p}}\right),
\label{Gamma-SP}
\end{equation}
where the zero-field (vacuum) result is
\begin{equation}
\Gamma_{\mathbf{p}}^{\text{SP}}(0) = \frac{\alpha_e^2 \lambda^2 m_\pi^4}{64 \pi^3 m_P^2 \sqrt{m_\pi^2+\mathbf{p}^2}}
= \frac{\alpha_e^2 m_\pi^4}{64 \pi^3 f_\pi^2 \sqrt{m_\pi^2+\mathbf{p}^2}}, 
\end{equation}
and the additional function appearing in the quadratic terms is defined through the integral
\begin{equation}
\mathcal{A}_{\mathbf{p}} =\frac{1}{2m_\pi} \int \frac{d^3 Q}{(2\pi)^3}\frac{1}{2 q_0}\int \frac{d^3 K}{(2\pi)^3}\frac{1}{2 k_0}(\mathbf{q}_\perp^2+\mathbf{q}_\perp\cdot\mathbf{k}_\perp+\mathbf{k}_{\perp}^2) (2\pi)^4\delta^{(4)}(p-k-q) .
\end{equation}
Since the integrand depends differently on the transverse and longitudinal components of the momenta, the function $\mathcal{A}_{\mathbf{p}}$, and consequently the decay rate $\Gamma_{\mathbf{p}}^{\text{SP}}$, is anisotropic. To quantify this directional dependence, we introduce the angle $\beta$ between the pion momentum and the magnetic field. The dependence of the decay rate on $\beta$, reflecting this magnetic-field-induced anisotropy, is studied numerically in the next section.
 
Similar to the center-of-mass result in Eq.~(\ref{Gamma-CS}), the general-frame expression in Eq.~(\ref{Gamma-SP}) shows that the decay rate remains suppressed by the magnetic field. The main new feature is the emergence of an angular dependence through the function $\mathcal{A}_{\mathbf{p}}$. As will be shown in the next section, this angular dependence leads to an enhanced suppression when the pion momentum is oriented perpendicular to the magnetic field.

\section{Numerical results}
\label{sec:numerical-results}

In this section, we present numerical results for the pion decay rate in a general reference frame, as given by Eq.~(\ref{Gamma-SP}). In this case, the pion carries a nonzero momentum $\mathbf{p}$, whose orientation relative to the magnetic field is specified by the angle $\beta$. We analyze how the decay rate depends on both the magnitude of $\mathbf{p}$ and its direction with respect to the field.

\begin{figure}
\begin{center}
\includegraphics[width=0.49\textwidth]{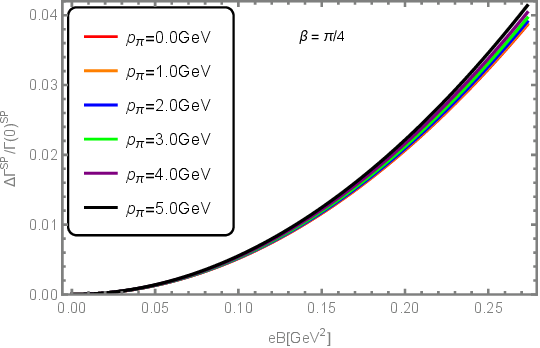}
\includegraphics[width=0.49\textwidth]{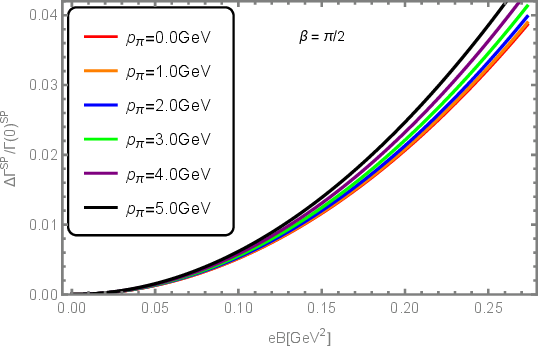}
\caption{Field-strength dependence of the absolute value of the field-induced correction to the pion decay rate in a general frame. The different curves correspond to various pion momenta, ranging from $p_\pi = 0$ to $p_\pi = 5~\text{GeV}$. Results are shown for two fixed angles: $\beta=\pi/4$ (left) and $\beta=\pi/2$ (right).}
\label{fig-rate-eB}
\end{center}
\end{figure}

Because the overall decay rate exhibits only a weak suppression by the magnetic field, it is useful to isolate the field-induced contribution by subtracting the zero-field result. To this end, we introduce the following dimensionless quantity:
\begin{equation}
\frac{|\Delta \Gamma_{\mathbf{p}}^{\text{SP}} (B)|}{\Gamma_{\mathbf{p}}^{\text{SP}} (0)}
= \frac{|\Gamma_{\mathbf{p}}^{\text{SP}} (B)-\Gamma_{\mathbf{p}}^{\text{SP}} (0)|}{\Gamma_{\mathbf{p}}^{\text{SP}} (0)}.
\label{Del-Gamma-SP}
\end{equation}
Its dependence on the magnetic-field strength is shown in Fig.~\ref{fig-rate-eB} for several values of the pion momentum, ranging from $p_\pi = 0$ to $p_\pi = 5~\text{GeV}$. The two panels correspond to fixed angles $\beta$. As seen in both panels, the magnitude of the field-dependent suppression increases quadratically with the magnetic-field strength, becoming more pronounced for larger pion momenta and for the orientations of pion momenta closer to $\beta = \pi/2$. Note that, for $|eB| \simeq m_\pi^2 \approx 0.018~\mbox{GeV}^2$, which is an optimistic benchmark for a late-time magnetic-field value, the correction remains extremely small, on the order of $0.02\%$. 

\begin{figure}
\begin{center}
\includegraphics[width=0.49\textwidth]{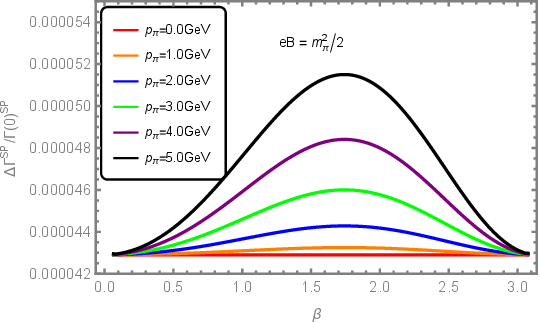}
\includegraphics[width=0.49\textwidth]{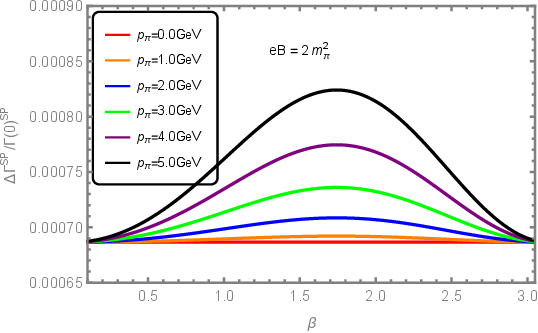}
\caption{Absolute value of the field-induced correction to the pion decay rate in a general frame as a function of the angle between the pion momentum and the magnetic field. The different curves correspond to various pion momenta, ranging from $p_\pi = 0$ to $p_\pi = 5~\text{GeV}$. Results are shown for two fixed magnetic-field strengths: $|eB|=m_\pi^2/2$ (left) and $|eB|=2m_\pi^2$ (right).}
\label{fig-rate-beta}
\end{center}
\end{figure}

The dependence of the field-induced correction to the decay rate given by Eq.~(\ref{Del-Gamma-SP}) on the angle $\beta$ is shown in Fig.~\ref{fig-rate-beta}. As seen from the figure, the absolute value of the field-induced correction reaches its maximum at $\beta = \pi/2$. This indicates that the suppression of the decay rate by the magnetic field is strongest when the pion momentum is perpendicular to the field direction. Although the quantitative effect remains small, it becomes more pronounced as the pion momentum increases.

It is important to emphasize, however, that this anisotropy remains extremely small in absolute magnitude. Therefore, the angular dependence identified here should be interpreted as a theoretical proof of principle. A quantitative connection between the small modification of the pion decay rate and an observable anisotropy of photon yields would require a realistic dynamical model for the time dependence of the magnetic field, the spacetime distribution of pion production and decay, and the subtraction of other photon sources. Such an analysis is beyond the scope of the present work.

\section{Discussion and Summary}
\label{sec:summary}
 
In this work, we investigated the decay of neutral pions into two photons in the presence of a weak background magnetic field within a hadronic framework. The analysis was based on the proton triangle diagram with an effective Yukawa-type $\pi^0$-proton coupling, which provides a physically transparent alternative to quark-level treatments in the regime where the magnetic length is much larger than the proton size. By expanding the proton propagator and the Schwinger phases in powers of $eB$, we derived analytical expressions for the decay amplitude and rate up to second order in the magnetic-field strength.
 
Our results show that the magnetic field leads to a generic suppression of the $\pi^0 \to \gamma\gamma$ decay rate. The effect originates from field-induced modifications of the intermediate proton loop and manifests as a quadratic dependence on the magnetic field, scaling as $|eB|^2/m_P^4$, where $m_P$ is the proton mass. Quantitatively, the correction is extremely small: for magnetic fields of order $|eB| \simeq m_\pi^2 \approx 0.018~\text{GeV}^2$, the suppression amounts to roughly $0.02\%$, increasing to only a few percent for fields an order of magnitude stronger. These values should be viewed as benchmark estimates. In realistic heavy-ion collisions, the magnetic field during the pion-rich stage is likely weaker than the values considered here, making the corresponding effect on the neutral-pion decay rate even smaller.

Although the field-induced suppression of the decay rate is qualitatively consistent with previous studies based on different theoretical frameworks~\cite{Shushpanov:1997sf,Agasian:2001ym,Simonov:2015xta,Coppola:2025nus,Andersen:2012zc,Andersen:2012dz,Brauner:2017uiu}, our quantitative results differ. The precise origin of these differences is not yet fully understood, but they likely arise from distinct model assumptions (e.g., the treatment of the pion as a pointlike particle) and from differences in the organization of the expansion in powers of the magnetic field.

In a general reference frame, the decay rate becomes anisotropic due to the different dependence of the amplitude on the transverse and longitudinal components of the pion momentum. This anisotropy is characterized by the angle $\beta$ between the pion momentum and the magnetic-field direction. The suppression reaches its maximum at $\beta = \pi/2$, when the pion momentum is perpendicular to the field, while it diminishes as the momentum becomes more aligned with the field.

Overall, our study should be interpreted primarily as a proof-of-principle demonstration that a weak background magnetic field can induce a directiospacetimen-dependent correction to the $\pi^0\to\gamma\gamma$ decay amplitude and rate. Quantitatively, however, the effect appears to be extremely small. In realistic heavy-ion collisions, the magnetic field during the pion-rich stage may be weaker than the benchmark values considered here. Then, the resulting anisotropy would be too small to be measurable with present experimental capabilities.

A quantitative estimate of how this small modification of the pion decay rate propagates into observable photon-yield anisotropies is highly model dependent and lies beyond the scope of the present work, since it would require coupling the decay calculation to a realistic description of the spacetime evolution of the magnetic field, pion production, and competing photon sources. Even if the photon sector were to inherit an anisotropy of comparable magnitude, such a signal would likely remain below current experimental uncertainties. Future work could address this question within a more complete phenomenological framework.

\begin{acknowledgments}
X.~W. would thank Mei Huang, Kun Xu for early contributions. The work of X.~W. was supported by YJ20240001 from Anhui University of Science and Technology. The work of I.~S. was supported in part by the U.S. National Science Foundation under Grants No.~PHY-2209470 and No.~PHY-2514933.
\end{acknowledgments}

\appendix

\section{Fermion propagator in a magnetic field}
\label{A1}
In a constant background field pointing in the $z$ direction, Green's function of a fermion with four-momentum $k$ and mass $m$ is given by~\cite{Miransky:2015ava}
\begin{equation}
G(x,x')=\Phi(x,x')\int \frac{d^4k}{(2\pi)^4}e^{-ik\cdot(x-x')}\hat{G}(k),
\end{equation}
where the Schwinger phase is given by
\begin{equation}
\Phi(x,x')=\exp\left\{i|e|\int_{x'}^{x}d\xi^\mu\left[A_{\mu}+\frac{1}{2}F_{\mu\nu}(\xi-x')^{\nu}\right]\right\}.
\label{Schwinger-phase}
\end{equation}
The translation-invariant part is given by
\begin{eqnarray}
\label{Gh}
\hat{G}(k) = i e^{-k_{\perp}^2 l^2} \sum_{n=0}^{\infty}\frac{(-1)^n D_{n}(k) }{k_0^2 -k_3^2-m^2-2n|e B| },
\end{eqnarray}
where 
\begin{equation}
D_n(k) = 2[k^0 \gamma^0 - k^3 \gamma^3 +  m][\mathcal{P}_{+}L_{n}(2\mathbf{k}^{2}_{\perp}l^2)- \mathcal{P}_{-}L_{n-1}(2\mathbf{k}^{2}_{\perp}l^2)]+4(\mathbf{k}_{\perp}\cdot\bm{\gamma}_{\perp})L_{n-1}^{1}(2k^{2}_{\perp}l^2).
\end{equation}

\subsection{Weak-field limit of the translation-invariant part of the propagator}

To obtain the weak-field approximation for the translation-invariant part of the propagator, it is convenient to use the following relation:
\begin{equation}
\frac{1}{a+2n|b|}=-\int_{0}^{\infty} ds e^{-s(a+2n|b|)},
\end{equation}
where we used the shorthand notations $a =(m^2+k_3^2-k_0^2)$ and $b = e B$. Then, we derive~\cite{Gorbar:2013uga}
\begin{eqnarray}
I_1& =& 2e^{-k_{\perp}^2 l^2}\sum_{n=0}^{\infty}\frac{(-1)^n[\mathcal{P}_{+}L_{n}(2\mathbf{k}^{2}_{\perp}l^2)- \mathcal{P}_{-}L_{n-1}(2\mathbf{k}^{2}_{\perp}l^2)]}{a+2n|b|}\nonumber\\
&=& -2 e^{-k_{\perp}^2/|b|}\int_0^{\infty}~ds\sum_{n=0}^{\infty}(-1)^n e^{-s(a+2n|b|)}\left[\mathcal{P}_{-}+\mathcal{P}_{+}e^{-2s|b|}\right]L_{n}\left(\frac{2\mathbf{k}_{\perp}^2}{|b|}\right) \nonumber\\
&=&-\int_{0}^{\infty}ds e^{-sa-(\mathbf{k}_{\perp}^2/b)\tanh(sb)}\left[ 1-i \gamma^1 \gamma^2 \tanh(sb)\right] \nonumber\\
&\simeq&-\frac{1}{a+\mathbf{k}_{\perp}^2}+i\gamma^1\gamma^2 \frac{1}{(a+\mathbf{k}_{\perp}^2)^2}b -\frac{2 \mathbf{k}_{\perp}^2}{(a+\mathbf{k}_{\perp}^{2})^4}b^2+O(b^3).
\label{app-I1}
\end{eqnarray}
Similarly, 
\begin{eqnarray}
I_2& =&4e^{-k_{\perp}^2 l^2}\sum_{n=0}^{\infty}\frac{(-1)^n L_{n-1}^{1}(2\mathbf{k}^{2}_{\perp}l^2)}{a+2n|b|}
= 4 e^{-\mathbf{k}_{\perp}^2/|b|}\int_0^{\infty}~ds\sum_{n=0}^{\infty}(-1)^n e^{-s(a+2n|b|+2|b|)}L_{n}^1\left(\frac{2\mathbf{k}_{\perp}^2}{|b|}\right)\nonumber\\
&=&\int_{0}^{\infty}\frac{ds}{\cosh^{2}(sb)} e^{-s a -(\mathbf{k}_{\perp}^2/b)\tanh(sb)}
\simeq \frac{1}{a+\mathbf{k}_{\perp}^2}+\frac{2 \mathbf{k}_{\perp}^2}{(a+\mathbf{k}_{\perp}^{2})^4}b^2+O(b^3).
\label{app-I2}
\end{eqnarray}
In the derivation, we utilized the definition of the generating function for the Laguerre polynomials~\cite{Gradshtein:1980},
\begin{equation}
\sum_{n=0}^{\infty}z^{n}L_{n}^{\alpha}(x)=\frac{1}{(1-z)^{1+\alpha}}\exp\left(\frac{x z}{z-1}\right),
\end{equation}
and took into account our convention $L_{-1}^{\alpha}(x) = 0$. 

By combining the results in Eqs.~(\ref{app-I1}) and (\ref{app-I2}), we obtain the following power series expression for the translation-invariant part of the propagator in weak-field limit:
\begin{equation}
\hat{G}(k_{\perp}, k_{\parallel}) = \hat{G}^{(0)}(k_{\perp}, k_{\parallel})+\hat{G}^{(1)}(k_{\perp}, k_{\parallel})+\hat{G}^{(2)}(k_{\perp}, k_{\parallel})+ \cdots,
\end{equation}
where the explicit expressions for the first three terms in the expansion are
\begin{subequations}
\label{prop-weak}
\begin{eqnarray}
 \hat{G}^{(0)}(k_{\perp}, k_{\parallel}) &=& i\frac{k_0\gamma^0-\mathbf{k}\cdot\bm{\gamma}+m}{k_0^2-\mathbf{k}^2-m^2},
\\
 \hat{G}^{(1)}(k_{\perp}, k_{\parallel}) &=& -\gamma^1 \gamma^2 eB\frac{k_0\gamma^0-k^3 \gamma^3+m}{(k_0^2-\mathbf{k}^2-m^2)^2},
\\
 \hat{G}^{(2)}(k_{\perp}, k_{\parallel}) &=& 2i(e B)^2\mathbf{k}_{\perp}^{2}\frac{k_0\gamma^0-\mathbf{k}\cdot\bm{\gamma}+m}{(k_0^2-\mathbf{k}^2-m^2)^4}.
 \end{eqnarray}
\end{subequations}

\subsection{Weak-field limit of the product of the Schwinger phase factors}

In the calculation of the triangle diagram, we encounter the following product of three Schwinger phases:
\begin{equation}
\Phi(x,z)\Phi(z,y)\Phi(y,x) = e^{\frac{i|e B|}{2}\epsilon^{ij}(y-x)_{i}(x-z)_{j}}.
\end{equation}
By using the weak-field limit, the corresponding expression can be written as a power series, 
\begin{equation}
\Phi(x,z)\Phi(z,y)\Phi(y,x) \simeq  1 +\frac{i |e B|\epsilon^{ij}(y-x)_{i}(x-z)_{j}}{2}
-\frac{|e B|^2\left[\epsilon^{ij}(y-x)_{i}(x-z)_{j}\right]^2}{4}+O(|eB|^3),
\end{equation}
where the indices $i,j =1,2$ and $\epsilon^{ij}$ is the Levi-Civita symbol. Therefore, in the momentum space, we have
\begin{eqnarray}
I_3 &=& \int d^4 x d^4 y d^4 z e^{-ir\cdot (z-x)}e^{-is\cdot (y-z)}e^{-it\cdot (x-y)}e^{-ip \cdot x}e^{i k\cdot y} e^{i q \cdot z}\Phi(x,z)\Phi(z,y)\Phi(y,x)\nonumber\\
&\simeq&(2\pi)^4\delta^{(4)}(p-k-q)\int d^4 y d^4 z e^{i z(s+q-r) -iy(s-t-k)}\left(1- \frac{i |e B|\epsilon^{ij}y_{i} z_{j}}{2}-\frac{|e B|^2\left(\epsilon^{ij}y_{i}z_{j}\right)^2}{4}\right).
\end{eqnarray}
Integrating over the spatial coordinate, we finally obtain 
\begin{equation}
I_3 =(2\pi)^4\delta^{(4)}(p-k-q) \left( \mathcal{I}_0+\mathcal{I}_1+\mathcal{I}_2 \right),
\end{equation}
where 
\begin{subequations}
\label{phase-weak}
\begin{equation}
\mathcal{I}_0 =(2\pi)^8\delta^{(4)}(s+q-r)\delta^{(4)}(s-t-k),
\end{equation}
\begin{eqnarray}
\mathcal{I}_1& =&\frac{i |e B|}{2}(2\pi)^8\delta^{(2)}(s_{\parallel}+q_{\parallel}-r_{\parallel})\delta^{(2)}(s_{\parallel}-t_{\parallel}-k_{\parallel})\left[-\delta(s_y-t_y-k_y)\delta'(s_x-t_x-k_x)\right.\nonumber\\
&\times&\left.\delta(s_x+q_x-r_x)\delta'(s_y+q_y-r_y)+\delta(s_x-t_x-k_x)\delta'(s_y-t_y-k_y)\delta(s_y+q_y-r_y)\delta'(s_x+q_x-r_x) \right],
\end{eqnarray}
\begin{eqnarray}
\mathcal{I}_2&=&\frac{|e B|^2}{4}(2\pi)^8\delta^{(2)}(s_{\parallel}+q_{\parallel}-r_{\parallel})\delta^{(2)}(s_{\parallel}-t_{\parallel}-k_{\parallel})
\left[-\delta(s_y-t_y-k_y)\delta''(s_x-t_x-k_x)\delta(s_x+q_x-r_x)\delta''(s_y+q_y-r_y)\right.\nonumber\\
&-&\delta''(s_y-t_y-k_y)\delta(s_x-t_x-k_x)\delta(s_y+q_y-r_y)\delta''(s_x+q_x-r_x)\nonumber\\
& +&\left.2\delta'(s_y-t_y-k_y)\delta'(s_x-t_x-k_x)\delta'(s_y+q_y-r_y)\delta'(s_x+q_x-r_x)\right].
\end{eqnarray}
\end{subequations}

\section{Evaluation of $\mathcal{F}_{\kappa}^{(j)}$ functions}
\label{App-Func}

By using the expansions for the translation-invariant part of the propagator and the product of the Schwinger phases in Eq.~(\ref{prod-three-phases}), we derive the following expression for the amplitude for the pion decay:
\begin{eqnarray}
M^{\mu \nu}_{1} & \simeq&\int \frac{d^4 r}{(2\pi)^4} \frac{d^4 s}{(2\pi)^4} \frac{d^4 t}{(2\pi)^4}\left(\mathcal{I}_0+\mathcal{I}_1+\mathcal{I}_2\right) \text{Tr} \Bigg[ \gamma^{\mu}\left(\hat{G}^{(0)}(s)+\hat{G}^{(1)}(s)+\hat{G}^{(2)}(s)\right)\gamma^{\nu}\left(\hat{G}^{(0)}(r)+\hat{G}^{(1)}(r)+\hat{G}^{(1)}(r)\right)\nonumber\\
& \times& \gamma^{5}\left(\hat{G}^{(0)}(t)+\hat{G}^{(1)}(t)+\hat{G}^{(1)}(t)\right)\Bigg]
\simeq\mathcal{F}_0^{(0)}+\sum_{j=0}^{3}\mathcal{F}_1^{(j)}+\sum_{j=0}^{9}\mathcal{F}_2^{(j)},
\end{eqnarray}
with the functions $\hat{G}^{(i)}(s)$ and $\mathcal{I}_i$ ($i=0,1,2$) defined in Eqs.~(\ref{prop-weak})
 and (\ref{phase-weak}), respectively. The explicit expressions for partial contributions $\mathcal{F}_{\kappa}^{(j)}$ read
\begin{subequations}
\begin{eqnarray}
\mathcal{F}_0^{(0)}=\int \frac{d^4 r}{(2\pi)^4} \frac{d^4 s}{(2\pi)^4} \frac{d^4 t}{(2\pi)^4}\mathcal{I}_0 \text{Tr} \left [ \gamma^{\mu}\hat{G}^{(0)}(s)\gamma^{\nu}\hat{G}^{(0)}(r)\gamma^{5}\hat{G}^{(0)}(t)\right],
\\
\mathcal{F}_1^{(0)}=\int \frac{d^4 r}{(2\pi)^4} \frac{d^4 s}{(2\pi)^4} \frac{d^4 t}{(2\pi)^4}\mathcal{I}_1 \text{Tr} \left [ \gamma^{\mu}\hat{G}^{(0)}(s)\gamma^{\nu}\hat{G}^{(0)}(r)\gamma^{5}\hat{G}^{(0)}(t)\right],
\\
\mathcal{F}_1^{(1)}=\int \frac{d^4 r}{(2\pi)^4} \frac{d^4 s}{(2\pi)^4} \frac{d^4 t}{(2\pi)^4}\mathcal{I}_0 \text{Tr} \left [ \gamma^{\mu}\hat{G}^{(1)}(s)\gamma^{\nu}\hat{G}^{(0)}(r)\gamma^{5}\hat{G}^{(0)}(t)\right],
\\
\mathcal{F}_1^{(2)}=\int \frac{d^4 r}{(2\pi)^4} \frac{d^4 s}{(2\pi)^4} \frac{d^4 t}{(2\pi)^4}\mathcal{I}_0 \text{Tr} \left [ \gamma^{\mu}\hat{G}^{(0)}(s)\gamma^{\nu}\hat{G}^{(1)}(r)\gamma^{5}\hat{G}^{(0)}(t)\right],
\\
\mathcal{F}_1^{(3)}=\int \frac{d^4 r}{(2\pi)^4} \frac{d^4 s}{(2\pi)^4} \frac{d^4 t}{(2\pi)^4}\mathcal{I}_0 \text{Tr} \left [ \gamma^{\mu}\hat{G}^{(0)}(s)\gamma^{\nu}\hat{G}^{(0)}(r)\gamma^{5}\hat{G}^{(1)}(t)\right],
\\
\mathcal{F}_2^{(0)}=\int \frac{d^4 r}{(2\pi)^4} \frac{d^4 s}{(2\pi)^4} \frac{d^4 t}{(2\pi)^4}\mathcal{I}_2 \text{Tr} \left [ \gamma^{\mu}\hat{G}^{(0)}(s)\gamma^{\nu}\hat{G}^{(0)}(r)\gamma^{5}\hat{G}^{(0)}(t)\right].
\\
\mathcal{F}_2^{(1)}=\int \frac{d^4 r}{(2\pi)^4} \frac{d^4 s}{(2\pi)^4} \frac{d^4 t}{(2\pi)^4}\mathcal{I}_0 \text{Tr} \left [ \gamma^{\mu}\hat{G}^{(2)}(s)\gamma^{\nu}\hat{G}^{(0)}(r)\gamma^{5}\hat{G}^{(0)}(t)\right],
\\
\mathcal{F}_2^{(2)}=\int \frac{d^4 r}{(2\pi)^4} \frac{d^4 s}{(2\pi)^4} \frac{d^4 t}{(2\pi)^4}\mathcal{I}_0 \text{Tr} \left [ \gamma^{\mu}\hat{G}^{(0)}(s)\gamma^{\nu}\hat{G}^{(2)}(r)\gamma^{5}\hat{G}^{(0)}(t)\right],
\\
\mathcal{F}_2^{(3)}=\int \frac{d^4 r}{(2\pi)^4} \frac{d^4 s}{(2\pi)^4} \frac{d^4 t}{(2\pi)^4}\mathcal{I}_0 \text{Tr} \left [ \gamma^{\mu}\hat{G}^{(0)}(s)\gamma^{\nu}\hat{G}^{(0)}(r)\gamma^{5}\hat{G}^{(2)}(t)\right],
\\
\mathcal{F}_2^{(4)}=\int \frac{d^4 r}{(2\pi)^4} \frac{d^4 s}{(2\pi)^4} \frac{d^4 t}{(2\pi)^4}\mathcal{I}_0 \text{Tr} \left [ \gamma^{\mu}\hat{G}^{(1)}(s)\gamma^{\nu}\hat{G}^{(1)}(r)\gamma^{5}\hat{G}^{(0)}(t)\right],
\\
\mathcal{F}_2^{(5)}=\int \frac{d^4 r}{(2\pi)^4} \frac{d^4 s}{(2\pi)^4} \frac{d^4 t}{(2\pi)^4}\mathcal{I}_0 \text{Tr} \left [ \gamma^{\mu}\hat{G}^{(1)}(s)\gamma^{\nu}\hat{G}^{(0)}(r)\gamma^{5}\hat{G}^{(1)}(t)\right],
\\
\mathcal{F}_2^{(6)}=\int \frac{d^4 r}{(2\pi)^4} \frac{d^4 s}{(2\pi)^4} \frac{d^4 t}{(2\pi)^4}\mathcal{I}_0 \text{Tr} \left [ \gamma^{\mu}\hat{G}^{(0)}(s)\gamma^{\nu}\hat{G}^{(1)}(r)\gamma^{5}\hat{G}^{(1)}(t)\right],
\\
\mathcal{F}_2^{(7)}=\int \frac{d^4 r}{(2\pi)^4} \frac{d^4 s}{(2\pi)^4} \frac{d^4 t}{(2\pi)^4}\mathcal{I}_1 \text{Tr} \left [ \gamma^{\mu}\hat{G}^{(1)}(s)\gamma^{\nu}\hat{G}^{(0)}(r)\gamma^{5}\hat{G}^{(0)}(t)\right],
\\
\mathcal{F}_2^{(8)}=\int \frac{d^4 r}{(2\pi)^4} \frac{d^4 s}{(2\pi)^4} \frac{d^4 t}{(2\pi)^4}\mathcal{I}_1 \text{Tr} \left [ \gamma^{\mu}\hat{G}^{(0)}(s)\gamma^{\nu}\hat{G}^{(1)}(r)\gamma^{5}\hat{G}^{(0)}(t)\right],
\\
\mathcal{F}_2^{(9)}=\int \frac{d^4 r}{(2\pi)^4} \frac{d^4 s}{(2\pi)^4} \frac{d^4 t}{(2\pi)^4}\mathcal{I}_1 \text{Tr} \left [ \gamma^{\mu}\hat{G}^{(0)}(s)\gamma^{\nu}\hat{G}^{(0)}(r)\gamma^{5}\hat{G}^{(1)}(t)\right],
\end{eqnarray}
\end{subequations}

\subsection{Zeroth-order contributions}

The zeroth-order contribution is only from the $\mathcal{F}_0^{(0)}$ term,
\begin{eqnarray}
\label{F00}
\mathcal{F}_0^{(0)}&=&\int \frac{d^4 r}{(2\pi)^4} \frac{d^4 s}{(2\pi)^4} \frac{d^4 t}{(2\pi)^4}\mathcal{I}_0 \text{Tr} \left [ \gamma^{\mu}\hat{G}^{(0)}(s)\gamma^{\nu}\hat{G}^{(0)}(r)\gamma^{5}\hat{G}^{(0)}(t)\right]\nonumber\\
& =&4 m_P \epsilon^{\mu\nu\alpha\beta} k_{\alpha} q_{\beta}\int\frac{d^4 s}{(2\pi)^4}  \frac{1}{(s^2-m_P^2)[(s+q)^2-m_P^2][(s-k)^2-m_P^2]}\nonumber\\
&=&4 m_P \epsilon^{\mu \nu \alpha \beta} k_{\alpha} q_{\beta} \left(\frac{-i}{16\pi^2}\right)\int^{1}_{0}dx \int_{0}^{1-x} dy \frac{1}{m_P^2-x(1-x)k^2-y(1-y)q^2-2 k q xy}.
\end{eqnarray}
Up to here, it is clearly shown that $\mathcal{F}_0^{(0)} = \mathcal{F}_0^{(0)}(q\leftrightarrow k, \mu\leftrightarrow \nu )$; therefore, 
\begin{equation}
\mathcal{F}_0^{\text{All}} = \mathcal{F}_0^{(0)} + \mathcal{F}_0^{(0)}(q\leftrightarrow k, \mu\leftrightarrow \nu ) = 2 \mathcal{F}_0^{(0)}.
\end{equation}
After use of the on-shell conditions, Eq.~(\ref{F00}) becomes
\begin{eqnarray}
\label{F00-F}
\mathcal{F}_0^{(0)}&=&4 m_P \epsilon^{\mu \nu \alpha \beta} k_{\alpha} q_{\beta} \left(\frac{-i}{16\pi^2}\right)\int^{1}_{0}dx \int_{0}^{1-x} dy \frac{1}{m_P^2-m_\pi^2 xy}\nonumber\\
&=&4 m_P \epsilon^{\mu \nu \alpha \beta} k_{\alpha} q_{\beta} \left(\frac{-i}{16\pi^2 m_\pi^2}\right)\left[\text{Li}_2\left(\frac{2 m_\pi}{m_\pi-\sqrt{m_\pi^2-4 m_P^2}}\right)+\text{Li}_2\left(\frac{2 m_\pi}{m_\pi+\sqrt{m_\pi^2-4 m_P^2}}\right)\right].
\end{eqnarray}
By using the small $m_\pi$ limit, we derive 
\begin{equation}
\lim_{m_\pi\rightarrow0}\frac{1}{m_\pi^2} 
\left[\text{Li}_2\left(\frac{2 m_\pi}{m_\pi-\sqrt{m_\pi^2-4 m_P^2}}\right)+\text{Li}_2\left(\frac{2 m_\pi}{m_\pi+\sqrt{m_\pi^2-4 m_P^2}}\right)\right]
= \frac{1}{2 m_P^2},
\end{equation}
leading to
\begin{equation}
\label{Final-Vaccum}
\mathcal{F}_0^{\text{All},sp} = -i \epsilon^{\mu \nu \alpha \beta} \frac{k_{\alpha} q_{\beta} }{4\pi^2 m_P},
\end{equation}
which agrees with the textbook result in Ref.~\cite{Schwartz:2014sze}.

\subsection{First-order contributions}

At the leading order in $e B$, we have four different contributions:
\begin{eqnarray}
\mathcal{F}_{1}^{(0)} &=& \frac{i|e B|}{2}\int \frac{d^4 s dr^2_{\perp}  dt^2_{\perp}}{(2\pi)^4}\left[-\delta(s_y-t_y-k_y)\delta'(s_x-t_x-k_x)\delta(s_x+q_x-r_x)\delta'(s_y+q_y-r_y)\right.\nonumber\\
&&+\left.\delta(s_x-t_x-k_x)\delta'(s_y-t_y-k_y)\delta(s_y+q_y-r_y)\delta'(s_x+q_x-r_x)\right]\nonumber\\
&&\times\text{Tr} \left [ \gamma^{\mu}\hat{G}^{(0)}(s)\gamma^{\nu}\hat{G}^{(0)}(s_{\parallel}+q_{\parallel},\mathbf{r}_{\perp})\gamma^{5}\hat{G}^{(0)}(s_{\parallel}-k_{\parallel},\mathbf{t}_{\perp})\right]\nonumber\\
&=& \frac{|e B|}{2}\int \frac{d^4 s }{(2\pi)^4}\text{Tr}\gamma^{\mu} \frac{(\slashed{s}+m_P)}{(s^2-m_P^2)}\gamma^{\nu}\nonumber\\
&\times&\left\{\frac{\gamma^1[-(s+q)^2+m_P^2]+2(s_x+q_x)[\slashed{s}+\slashed{q}+m_P]}{[(s+q)^2-m_P^2]^2}\gamma^5\frac{\gamma^2[-(s-k)^2+m_P^2]+2(s_y-k_y)[\slashed{s}-\slashed{k}+m_P]}{[(s-k)^2-m_P^2]^2}\right.\nonumber\\
&-& \left.\frac{\gamma^2[-(s+q)^2+m_P^2]+2(s_y+q_y)[\slashed{s}+\slashed{q}+m_P]}{[(s+q)^2-m_P^2]^2}\gamma^5\frac{\gamma^1[-(s-k)^2+m_P^2]+2(s_x-k_x)[\slashed{s}-\slashed{k}+m_P]}{[(s-k)^2-m_P^2]^2}\right\},
\label{F10}
\end{eqnarray}
\begin{eqnarray}
\label{F01}
\mathcal{F}_1^{(1)}&=& e B \int\frac{d^4 s}{(2\pi)^4}\left\{ \gamma^{\mu} \gamma^1\gamma^2\frac{\gamma_{\parallel}^{\sigma}k_{\parallel, \sigma}+m}{(s^2-m_P^2)^2}\gamma^{\nu} \frac{\gamma^{\rho}(s+q)_{\rho}+m_P}{[(s+q)^2-m_P^2]}\gamma^5\frac{ \gamma^{\tau}(s-q)_{\tau}+m_P}{[(s-k)^2-m_P^2]}\right\},
\\
\label{F02}
\mathcal{F}_1^{(2)}&=&  e B \int\frac{d^4 s}{(2\pi)^4}\left\{ \gamma^{\mu} \frac{\gamma^{\sigma}k_{\sigma}+m}{(s^2-m_P^2)}\gamma^{\nu} \gamma^1\gamma^2\frac{\gamma^{\rho}_{\parallel}(s+q)_{\parallel,\rho}+m_P}{[(s+q)^2-m_P^2]^2}\gamma^5\frac{ \gamma^{\tau}(s-k)_{\tau}+m_P}{[(s-k)^2-m_P^2]}\right\},
\\
\label{F03}
\mathcal{F}_1^{(3)}&=&  e B \int\frac{d^4 s}{(2\pi)^4}\left\{ \gamma^{\mu} \frac{\gamma^{\sigma}k_{\sigma}+m}{(s^2-m_P^2)}\gamma^{\nu} \frac{\gamma^{\rho}(s+q)_{\rho}+m_P}{[(s+q)^2-m_P^2]}\gamma^5\gamma^1\gamma^2\frac{ \gamma_{\parallel}^{\tau}(s-k)_{\parallel,\tau}+m_P}{[(s-k)^2-m_P^2]^2}\right\}.
\end{eqnarray}
In the first function, we have used 
\begin{equation}
\int_{-\infty}^{\infty}\delta^{(n)}(x)\phi(x)dx = \int_{-\infty}^{\infty}(-1)^n\delta(x)\phi^{(n)}(x) dx. 
\end{equation}
After tedious but straightforward calculations, we obtain 
\begin{eqnarray}
\mathcal{F}_1^{(0)}+\mathcal{F}_1^{(0)}(q\leftrightarrow k, \mu\leftrightarrow \nu )=0,\\
\mathcal{F}_1^{(1)}+\mathcal{F}_1^{(1)}(q\leftrightarrow k, \mu\leftrightarrow \nu )  =0,\\
\mathcal{F}_1^{(2)} +\mathcal{F}_1^{(3)}(q\leftrightarrow k, \mu\leftrightarrow \nu ) =0,\\
\mathcal{F}_1^{(3)}+\mathcal{F}_1^{(2)}(q\leftrightarrow k, \mu\leftrightarrow \nu )=0 .
\end{eqnarray}
Consequently, all terms of first order in the magnetic field vanish and do not affect the decay rate.

\subsection{Second-order contributions}

For the next leading-order contributions that are promotional to $(e B)^2$, the first three terms are given by 
\begin{eqnarray}
\label{F0200}
\mathcal{F}_2^{(1)}&=&\int \frac{d^4 r}{(2\pi)^4} \frac{d^4 s}{(2\pi)^4} \frac{d^4 t}{(2\pi)^4}\mathcal{I}_0 \text{Tr} \left [ \gamma^{\mu}\hat{G}^{(2)}(s)\gamma^{\nu}\hat{G}^{(0)}(r)\gamma^{5}\hat{G}^{(0)}(t)\right]\nonumber\\
& =&8 |e B|^2  m_P \epsilon^{\mu\nu\alpha\beta} k_{\alpha} q_{\beta}\int\frac{d^4 s}{(2\pi)^4}  \frac{\mathbf{s}_{\perp}^2}{(s^2-m_P^2)^4[(s+q)^2-m_P^2][(s-k)^2-m_P^2]},
\end{eqnarray}
\begin{eqnarray}
\label{F0020}
\mathcal{F}_2^{(2)}&=&\int \frac{d^4 r}{(2\pi)^4} \frac{d^4 s}{(2\pi)^4} \frac{d^4 t}{(2\pi)^4}\mathcal{I}_0 \text{Tr} \left [ \gamma^{\mu}\hat{G}^{(0)}(s)\gamma^{\nu}\hat{G}^{(2)}(r)\gamma^{5}\hat{G}^{(0)}(t)\right]\nonumber\\
  & =&8 |e B|^2  m_P \epsilon^{\mu\nu\alpha\beta} k_{\alpha} q_{\beta}\int\frac{d^4 s}{(2\pi)^4}  \frac{(\mathbf{s}_{\perp}+\mathbf{q}_{\perp})^2}{(s^2-m_P^2)[(s+q)^2-m_P^2]^4[(s-k)^2-m_P^2]},
\end{eqnarray}
\begin{eqnarray}
\label{F0002}
\mathcal{F}_2^{(3)}&=&\int \frac{d^4 r}{(2\pi)^4} \frac{d^4 s}{(2\pi)^4} \frac{d^4 t}{(2\pi)^4}\mathcal{I}_0 \text{Tr} \left [ \gamma^{\mu}\hat{G}^{(0)}(s)\gamma^{\nu}\hat{G}^{(0)}(r)\gamma^{5}\hat{G}^{(2)}(t)\right]\nonumber\\
  & =&8 |e B|^2  m_P \epsilon^{\mu\nu\alpha\beta} k_{\alpha} q_{\beta}\int\frac{d^4 s}{(2\pi)^4}  \frac{(\mathbf{s}_{\perp}-\mathbf{k}_{\perp})^2}{(s^2-m_P^2)[(s+q)^2-m_P^2][(s-k)^2-m_P^2]^4}.
\end{eqnarray}
We define $\mathcal{F}_{2,0}=\mathcal{F}_2^{(1)}+\mathcal{F}_2^{(2)}+\mathcal{F}_2^{(3)}$. Then, it is easy to show that
\begin{equation}
\label{F02A}
\mathcal{F}_{2,0}^{\text{All}} = \mathcal{F}_{2,0} +\mathcal{F}_{2,0}(q\leftrightarrow k, \mu\leftrightarrow\nu)= 2 \mathcal{F}_{2,0}.
\end{equation}
By using the results in Appendix~\ref{J-Cal}, we obtain  
\begin{equation}
\label{F20A-F}
\mathcal{F}_{2,0}^{\text{All}} = 16 |e B|^2  m_P \epsilon^{\mu\nu\alpha\beta} k_{\alpha} q_{\beta} \sum_{i=1}^{3} \left(\mathcal{J}_i^1 + \mathcal{J}_i^2\right),
\end{equation}
where $\sum_{i=1}^{3} \mathcal{J}_i^1$ and $\sum_{i=1}^{3} \mathcal{J}_i^2$ are given in Eq.~(\ref{I1_F}) and Eq.(\ref{I2_F}). Using the results in Eqs.~(\ref{Jsp}), (\ref{JCM}), and (\ref{JCS}), valid for the SP, the CM, and the CS approximation, respectively, we derive 
\begin{eqnarray}
\label{Final-F20}
\mathcal{F}_{2,0}^{\text{All},\text{CS}}&\simeq& 16 i|e B|^2  m_P \epsilon^{\mu\nu\alpha\beta} k_{\alpha} q_{\beta}\left( \frac{1}{320\pi^2 m_P^6}+\frac{ \mathbf{q}_\perp^2}{2016\pi^2 m_P^8} \right).
\end{eqnarray}
The other six terms at the next-leading order are given by
\begin{equation}
\label{F0110}
\mathcal{F}_2^{(4)}
=  i  (e B)^2\int\frac{d^4 s}{(2\pi)^4}\text{Tr}\left\{ \gamma^{\mu} \gamma^1\gamma^2\frac{s_0 \gamma^0 -s_z \gamma^3+m_P}{(s^2-m_P^2)^2}\gamma^{\nu}\gamma^1\gamma^2 \frac{(s_0+q_0)\gamma^0-(s_z+q_z)\gamma^3+m_P}{[(s+q)^2-m_P^2]^2}\gamma^5\frac{ \slashed{s}-\slashed{k}+m_P}{[(s-k)^2-m_P^2]}\right\}, 
\end{equation}
\begin{equation}
\label{F0101}
\mathcal{F}_2^{(5)} = i  (e B)^2\int\frac{d^4 s}{(2\pi)^4}\text{Tr}\left\{ \gamma^{\mu} \gamma^1\gamma^2\frac{s_0 \gamma^0 -s_z \gamma^3+m_P}{(s^2-m_P^2)^2}\gamma^{\nu} \frac{\slashed{s}+\slashed{q}+m_P}{[(s+q)^2-m_P^2]}\gamma^5\gamma^1\gamma^2 \frac{(s_0-k_0)\gamma^0-(s_z-k_z)\gamma^3+m_P}{[(s-k)^2-m_P^2]^2}\right\},
\end{equation}
\begin{equation}
\label{F0011}
\mathcal{F}_2^{(6)} = i  (e B)^2\int\frac{d^4 s}{(2\pi)^4}\text{Tr}\left\{ \gamma^{\mu} \frac{\slashed{s}+m_P}{(s^2-m_P^2)}\gamma^{\nu}\gamma^1\gamma^2 \frac{(s_0+q_0)\gamma^0-(s_z+q_z)\gamma^3+m_P}{[(s+q)^2-m_P^2]^2}\gamma^5\gamma^1\gamma^2 \frac{(s_0-k_0)\gamma^0-(s_z-k_z)\gamma^3+m_P}{[(s-k)^2-m_P^2]^2}\right\}, 
\end{equation}
\begin{eqnarray}
\mathcal{F}_{2}^{(7)} &=&  \frac{i|e B|}{2}\int \frac{d^4 s dr^2_{\perp}  dt^2_{\perp}}{(2\pi)^4}\Big[-\delta(s_y-t_y-k_y)\delta'(s_x-t_x-k_x)\delta(s_x+q_x-r_x)\delta'(s_y+q_y-r_y)\nonumber\\
&&+\delta(s_x-t_x-k_x)\delta'(s_y-t_y-k_y)\delta(s_y+q_y-r_y)\delta'(s_x+q_x-r_x)\Big]\nonumber\\
&&\times\text{Tr} \left [ \gamma^{\mu}\hat{G}^{(1)}(s)\gamma^{\nu}\hat{G}^{(0)}(s_{\parallel}+q_{\parallel},\mathbf{r}_{\perp})\gamma^{5}\hat{G}^{(0)}(s_{\parallel}-k_{\parallel},\mathbf{t}_{\perp})\right]\nonumber\\
&=& i\frac{|e B|^2}{2}\int \frac{d^4 s }{(2\pi)^4}\text{Tr}\gamma^{\mu}\gamma^1 \gamma^2 \frac{(s^0\gamma^0-s_z \gamma^3 +m_P)}{(s^2-m_P^2)^2}\gamma^{\nu}\nonumber\\
&\times&\left\{\frac{\gamma^1[-(s+q)^2+m_P^2]+2(s_x+q_x)[\slashed{s}+\slashed{q}+m_P]}{[(s+q)^2-m_P^2]^2}\gamma^5\frac{\gamma^2[-(s-k)^2+m_P^2]+2(s_y-k_y)[\slashed{s}-\slashed{k}+m_P]}{[(s-k)^2-m_P^2]^2}\right.\nonumber\\
&-& \left.\frac{\gamma^2[-(s+q)^2+m_P^2]+2(s_y+q_y)[\slashed{s}+\slashed{q}+m_P]}{[(s+q)^2-m_P^2]^2}\gamma^5\frac{\gamma^1[-(s-k)^2+m_P^2]+2(s_x-k_x)[\slashed{s}-\slashed{k}+m_P]}{[(s-k)^2-m_P^2]^2}\right\},
\label{F1100}
\end{eqnarray}
\begin{eqnarray}
\mathcal{F}_{2}^{(8)} &=& \frac{i|e B|}{2}\int \frac{d^4 s dr^2_{\perp}  dt^2_{\perp}}{(2\pi)^4}\left[-\delta(s_y-t_y-k_y)\delta'(s_x-t_x-k_x)\delta(s_x+q_x-r_x)\delta'(s_y+q_y-r_y)\right.\nonumber\\
&&+\left.\delta(s_x-t_x-k_x)\delta'(s_y-t_y-k_y)\delta(s_y+q_y-r_y)\delta'(s_x+q_x-r_x)\right]\nonumber\\
&&\times\text{Tr} \left [ \gamma^{\mu}\hat{G}^{(0)}(s)\gamma^{\nu}\hat{G}^{(1)}(s_{\parallel}+q_{\parallel},\mathbf{r}_{\perp})\gamma^{5}\hat{G}^{(0)}(s_{\parallel}-k_{\parallel},\mathbf{t}_{\perp})\right]\nonumber\\
&=& i \frac{|e B|^2}{2}\int \frac{d^4 s }{(2\pi)^4}\text{Tr}\gamma^{\mu} \frac{(\slashed{s}+m_P)}{(s^2-m_P^2)}\gamma^{\nu}\nonumber\\
&\times&\left\{\frac{\gamma^1[-(s+q)^2+m_P^2]+2(s_x+q_x)[\slashed{s}+\slashed{q}+m_P]}{[(s+q)^2-m_P^2]^2}\gamma^5\frac{\gamma^2[-(s-k)^2+m_P^2]+2(s_y-k_y)[\slashed{s}-\slashed{k}+m_P]}{[(s-k)^2-m_P^2]^2}\right.\nonumber\\
&-& \left.\frac{\gamma^2[-(s+q)^2+m_P^2]+2(s_y+q_y)[\slashed{s}+\slashed{q}+m_P]}{[(s+q)^2-m_P^2]^2}\gamma^5\frac{\gamma^1[-(s-k)^2+m_P^2]+2(s_x-k_x)[\slashed{s}-\slashed{k}+m_P]}{[(s-k)^2-m_P^2]^2}\right\},
\label{F1010}
\end{eqnarray}
\begin{eqnarray}
\mathcal{F}_{2}^{(9)} &=& \frac{i|e B|}{2}\int \frac{d^4 s dr^2_{\perp}  dt^2_{\perp}}{(2\pi)^4}\left[-\delta(s_y-t_y-k_y)\delta'(s_x-t_x-k_x)\delta(s_x+q_x-r_x)\delta'(s_y+q_y-r_y)\right.\nonumber\\
&&+\left.\delta(s_x-t_x-k_x)\delta'(s_y-t_y-k_y)\delta(s_y+q_y-r_y)\delta'(s_x+q_x-r_x)\right]\nonumber\\
&&\times\text{Tr} \left [ \gamma^{\mu}\hat{G}^{(0)}(s)\gamma^{\nu}\hat{G}^{(0)}(s_{\parallel}+q_{\parallel},\mathbf{r}_{\perp})\gamma^{5}\hat{G}^{(1)}(s_{\parallel}-k_{\parallel},\mathbf{t}_{\perp})\right]\nonumber\\
&=& i \frac{|e B|^2}{2}\int \frac{d^4 s }{(2\pi)^4}\text{Tr}\gamma^{\mu} \frac{(\slashed{s}+m_P)}{(s^2-m_P^2)}\gamma^{\nu}\nonumber\\
&\times&\left\{\frac{\gamma^1[-(s+q)^2+m_P^2]+2(s_x+q_x)[\slashed{s}+\slashed{q}+m_P]}{[(s+q)^2-m_P^2]^2}\gamma^5\frac{\gamma^2[-(s-k)^2+m_P^2]+2(s_y-k_y)[\slashed{s}-\slashed{k}+m_P]}{[(s-k)^2-m_P^2]^2}\right.\nonumber\\
&-& \left.\frac{\gamma^2[-(s+q)^2+m_P^2]+2(s_y+q_y)[\slashed{s}+\slashed{q}+m_P]}{[(s+q)^2-m_P^2]^2}\gamma^5\frac{\gamma^1[-(s-k)^2+m_P^2]+2(s_x-k_x)[\slashed{s}-\slashed{k}+m_P]}{[(s-k)^2-m_P^2]^2}\right\}.
\label{F1001}
\end{eqnarray}
It is straightforward to check that all of them vanish, i.e.,  
\begin{eqnarray}
\mathcal{F}_2^{(4)}+\mathcal{F}_2^{(5)}(q\leftrightarrow k, \mu\leftrightarrow\nu)&=& 0, \nonumber\\
\mathcal{F}_2^{(5)}+\mathcal{F}_2^{(4)}(q\leftrightarrow k, \mu\leftrightarrow\nu)&=& 0 ,\nonumber\\
\mathcal{F}_2^{(6)}+\mathcal{F}_2^{(6)}(q\leftrightarrow k, \mu\leftrightarrow\nu)&=& 0 ,\nonumber\\
\mathcal{F}_2^{(7)}+\mathcal{F}_2^{(8)}(q\leftrightarrow k, \mu\leftrightarrow\nu)&=& 0,\nonumber\\
\mathcal{F}_2^{(8)}+\mathcal{F}_2^{(7)}(q\leftrightarrow k, \mu\leftrightarrow\nu)&=& 0,\nonumber\\
\mathcal{F}_2^{(9)}+\mathcal{F}_2^{(9)}(q\leftrightarrow k, \mu\leftrightarrow\nu)&=& 0.
\end{eqnarray}
The only remaining term is given by
\begin{eqnarray}
\label{F2000}
\mathcal{F}_{2}^{(0)} &=& \frac{|e B|^2}{4}\int \frac{d^4 s dr^2_{\perp}  dt^2_{\perp}}{(2\pi)^4}\Big[-\delta(s_y-t_y-k_y)\delta''(s_x-t_x-k_x)\delta(s_x+q_x-r_x)\delta''(s_y+q_y-r_y)\nonumber\\
&-&\delta''(s_y-t_y-k_y)\delta(s_x-t_x-k_x)\delta(s_y+q_y-r_y)\delta''(s_x+q_x-r_x)+2\delta'(s_y-t_y-k_y)\delta'(s_x-t_x-k_x)\nonumber\\
& \times&\delta'(s_y+q_y-r_y)\delta'(s_x+q_x-r_x)\Big]\text{Tr} \left [ \gamma^{\mu}\hat{G}^{(0)}(s)\gamma^{\nu}\hat{G}^{(0)}(s_{\parallel}+q_{\parallel},\mathbf{r}_{\perp})\gamma^{5}\hat{G}^{(0)}(s_{\parallel}-k_{\parallel},\mathbf{t}_{\perp})\right]\nonumber\\
&=&-i m_P  |e B|^2\left[\frac{\partial^4 }{\partial t_x^2\partial r_y^2}-\frac{2\partial^4}{\partial t_y \partial r_x \partial t_x \partial r_y}+\frac{\partial^4}{\partial t_y^2 \partial r_x^2}\right]\int\frac{d^4 s}{(2\pi)^4}  \frac{\epsilon^{\mu\nu\alpha\beta} k_{\alpha} q_{\beta}}{(s^2-m_P^2)[(s+q)^2-m_P^2][(s-k)^2-m_P^2]}.
\end{eqnarray}
After simplification, it can be written as follows:
\begin{equation}
\label{F2000-2}
\mathcal{F}_{2}^{(0)} =  -\frac{m_P  |e B|^2}{4(2\pi)^2}\left[\frac{\partial^4}{\partial k_x^2\partial q_y^2}-\frac{2\partial^4}{\partial k_y \partial q_x \partial k_x \partial q_y}+\frac{\partial^4}{\partial k_y^2 \partial q_x^2}\right] \int_{0}^{1} dz  \int_{0}^{1-z} dy  \frac{ \epsilon^{\mu\nu\alpha\beta}k_\alpha q_\beta}{ (yq - zk)^2 - y q^2 - z k^2 +m_P^2}.
\end{equation}
Therefore, we obtained 
\begin{equation}
\mathcal{F}_{0,2}^{\text{All}} = \mathcal{F}_{2}^{(0)}+\mathcal{F}_{2}^{(0)}(q\leftrightarrow k, \mu\leftrightarrow\nu)= 2\mathcal{F}_{2}^{(0)}.
\end{equation}
Furthermore, after carrying out straightforward, albeit tedious, evaluations of the integrals, we derive 
\begin{eqnarray}
\mathcal{F}_{2}^{(0)} &=& -\frac{m_P  |e B|^2}{4(2\pi)^2} \epsilon^{\mu\nu\alpha\beta}\left[k_{\parallel,\alpha} q_{\parallel,\beta}\mathcal{K}_0 +k_{\parallel,\alpha} \delta_{1,\beta}\mathcal{K}_{q_x}+k_{\parallel,\alpha} \delta_{2,\beta}\mathcal{K}_{q_y} +q_{\parallel,\beta} \delta_{1,\alpha}\mathcal{K}_{k_x} \right.\nonumber\\
&+& q_{\parallel,\beta} \delta_{2,\alpha}\mathcal{K}_{k_y}+\delta_{2,\alpha}\delta_{1,\beta}\mathcal{K}_{q_x k_y}+\left.\delta_{1,\alpha}\delta_{2,\beta}\mathcal{K}_{q_y k_x}\right],
\end{eqnarray}
with 
\begin{subequations}
\label{K_gen}
\begin{eqnarray}
\mathcal{K}_0& =&\frac{1}{3 m_P^4 m_\pi^7 (4 m_P^2 - m_\pi^2)^{3/2}}\left\{-8 m_\pi \sqrt{4 m_P^2-m_\pi^2}\left[q_y^2 \left(6k_x^2 \left(30 m_P^2-7 m_\pi^2\right)+30 m_P^4-13 m_P^2 m_\pi^2+m_\pi^4\right) \right.\right.\nonumber\\
&+&k_x^2 \left(30 m_P^4-13 m_P^2 m_\pi^2+m_\pi^4\right)+q_x \left(-12k_x k_y q_y \left(30 m_P^2-7 m_\pi^2\right)-6k_x m_P^2 \left(30 m_P^2-7 m_\pi^2\right)\right)\nonumber\\
&+&q_x^2 \left(6 k_y^2 \left(30 m_P^2-7 m_\pi^2\right)+30 m_P^4-13 m_P^2 m_\pi^2+m_\pi^4\right)+k_y^2 \left(30 m_P^4-13 m_P^2 m_\pi^2+m_\pi^4\right)+6 k_y m_P^2 q_y \left(7 m_\pi^2-30 m_P^2\right)\nonumber\\
&-&\left. 36 m_P^4 m_\pi^2+9 m_P^2 m_\pi^4\right]\nonumber\\
&+&48\arctan\left(\frac{m_\pi}{\sqrt{4m_P^2-m_\pi^2}}\right)\left[2 k_x^2 m_P^2 \left(10 m_P^4-6 m_P^2 m_\pi^2+m_\pi^4\right)+q_y^2 \left(4 k_x^2 \left(30 m_P^4-12 m_P^2 m_\pi^2+m_\pi^4\right)+20 m_P^6\right.\right.\nonumber\\
&-&\left. 12 m_P^4 m_\pi^2+2 m_P^2 m_\pi^4\right)+q_x \left(-8 k_x k_y q_y \left(30 m_P^4-12 m_P^2 m_\pi^2+m_\pi^4\right)-4 k_x m_P^2 \left(30 m_P^4-12 m_P^2 m_\pi^2+m_\pi^4\right)\right)\nonumber\\
&+& 2 k_y^2 m_P^2 \left(10 m_P^4-6 m_P^2 m_\pi^2+m_\pi^4\right)+q_x^2 \left(4 k_y^2 \left(30 m_P^4-12 m_P^2 m_\pi^2+m_\pi^4\right)+20 m_P^6-12 m_P^4 m_\pi^2+2 m_P^2 m_\pi^4\right)\nonumber\\
&-&\left.\left.4 k_y m_P^2 q_y \left(30 m_P^4-12 m_P^2 m_\pi^2+m_\pi^4\right)+24 m_P^6 m_\pi^2+10 m_P^4 m_\pi^4-m_P^2 m_\pi^6\right]\right\},
\end{eqnarray}
\begin{eqnarray}
\mathcal{K}_{q_x}& =&\frac{1}{3 m_P^4 m_\pi^7 (4 m_P^2 - m_\pi^2)^{3/2}}\left\{-8 m_\pi \sqrt{4 m_P^2-m_\pi^2}\left[\right.q_x \left(q_y^2 \left(6 k_x^2 \left(30 m_P^2-7 m_\pi^2\right)+30 m_P^4-13 m_P^2 m_\pi^2+m_\pi^4\right)\right.\right.\nonumber\\
&+&\left. k_x^2 \left(30 m_P^4-13 m_P^2 m_\pi^2+m_\pi^4\right)+k_y^2 \left(90 m_P^4-39 m_P^2 m_\pi^2+3 m_\pi^4\right)+12 k_y m_P^2 q_y \left(7 m_\pi^2-30 m_P^2\right)-72 m_P^4 m_\pi^2+18 m_P^2 m_\pi^4\right)\nonumber\\
&+&q_x^2 \left(k_x k_y q_y \left(84 m_\pi^2-360 m_P^2\right)+k_x \left(42 m_P^2 m_\pi^2-180 m_P^4\right)\right)+k_x k_y q_y \left(-60 m_P^4+26 m_P^2 m_\pi^2-2 m_\pi^4\right)\nonumber\\
&+&\left.k_x q_y^2 \left(180 m_P^4-42 m_P^2 m_\pi^2\right)+q_x^3 \left(k_y^2 \left(180 m_P^2-42 m_\pi^2\right)+30 m_P^4-13 m_P^2 m_\pi^2+m_\pi^4\right)\right]\nonumber\\
&+&96\arctan\left(\frac{m_\pi}{\sqrt{4m_P^2-m_\pi^2}}\right)\left[q_x \left(k_x^2 m_P^2 \left(10 m_P^4-6 m_P^2 m_\pi^2+m_\pi^4\right)+q_y^2 \left(2 k_x^2 \left(30 m_P^4-12 m_P^2 m_\pi^2+m_\pi^4\right)\right.\right.\right.\nonumber\\
&+&\left.10 m_P^6-6 m_P^4 m_\pi^2+m_P^2 m_\pi^4\right)+3 k_y^2 m_P^2 \left(10 m_P^4-6 m_P^2 m_\pi^2+m_\pi^4\right)-4 k_y m_P^2 q_y \left(30 m_P^4-12 m_P^2 m_\pi^2+m_\pi^4\right)\nonumber\\
&-&\left. 24 m_P^6 m_\pi^2+10 m_P^4 m_\pi^4-m_P^2 m_\pi^6\right)+q_x^2 \left(k_x k_y q_y \left(-120 m_P^4+48 m_P^2 m_\pi^2-4 m_\pi^4\right)+k_x \left(-60 m_P^6+24 m_P^4 m_\pi^2-2 m_P^2 m_\pi^4\right)\right)\nonumber\\
&+&k_x k_y q_y \left(-20 m_P^6+12 m_P^4 m_\pi^2-2 m_P^2 m_\pi^4\right)+k_x q_y^2 \left(60 m_P^6-24 m_P^4 m_\pi^2+2 m_P^2 m_\pi^4\right)\nonumber\\
&+&\left.\left. q_x^3 \left(2 k_y^2 \left(30 m_P^4-12 m_P^2 m_\pi^2+m_\pi^4\right)+10 m_P^6-6 m_P^4 m_\pi^2+m_P^2 m_\pi^4\right)\right]\right\},
\end{eqnarray}
\begin{eqnarray}
\mathcal{K}_{q_x k_y}& =&\frac{4}{3 m_P^4 m_\pi^7}\left\{\frac{m_\pi}{4m_P^6-m_P^4 m_\pi^2}\left[q_x \left(k_y \left(q_y^2 \left(-12 k_x^2 \left(30 m_P^2-7 m_\pi^2\right)-2 \left(30 m_P^4-13 m_P^2 m_\pi^2+m_\pi^4\right)\right)\right.\right.\right.\right.\nonumber\\
&+&\left.\left.\left. k_x^2 \left(-60 m_P^4+26 m_P^2 m_\pi^2-2 m_\pi^4\right)+354 m_P^4 m_\pi^2-90 m_P^2 m_\pi^4\right)+q_y \left(-12 k_x^2 m_P^2 \left(30 m_P^2-7 m_\pi^2\right)\right.\right.\right. \nonumber\\
&-&\left.\left.\left.  18 m_P^4 m_\pi^2+3 m_P^2 m_\pi^4\right)+k_y^3 \left(-180 m_P^4+78 m_P^2 m_\pi^2-6 m_\pi^4\right)+24 k_y^2 m_P^2 q_y \left(30 m_P^2-7 m_\pi^2\right)\right)\right.   \nonumber\\
&+& \left. q_x^2 \left(24 k_x k_y^2 q_y \left(30 m_P^2-7 m_\pi^2\right)+k_x k_y \left(720 m_P^4-168 m_P^2 m_\pi^2\right)+4 k_x q_y \left(30 m_P^4-13 m_P^2 m_\pi^2+m_\pi^4\right)\right)\right.   \nonumber\\
&+& \left.  k_x k_y^2 q_y \left(120 m_P^4-52 m_P^2 m_\pi^2+4 m_\pi^4\right)+k_y \left(k_x q_y^2 \left(84 m_P^2 m_\pi^2-360 m_P^4\right)+k_x \left(3 m_P^2 m_\pi^4-18 m_P^4 m_\pi^2\right)\right)
\right.   \nonumber\\
&+& \left.  k_x q_y \left(18 m_P^2 m_\pi^4-78 m_P^4 m_\pi^2\right)+q_x^3 \left(-12 k_y^3 \left(30 m_P^2-7 m_\pi^2\right)-6 k_y \left(30 m_P^4-13 m_P^2 m_\pi^2+m_\pi^4\right)\right)\right]\nonumber\\
&+&\frac{12}{m_P^4(4m_P^2-m_\pi^2)^{\frac{3}{2}}}\arctan\left(\frac{m_\pi}{\sqrt{4m_P^2-m_\pi^2}}\right)\left[q_x \left(k_y \left(2 k_x^2 m_P^2 \left(10 m_P^4-6 m_P^2 m_\pi^2+m_\pi^4\right)\right.\right.\right.\nonumber\\
&+&\left. q_y^2 \left(4 k_x^2 \left(30 m_P^4-12 m_P^2 m_\pi^2+m_\pi^4\right)+20 m_P^6-12 m_P^4 m_\pi^2+2 m_P^2 m_\pi^4\right)-110 m_P^6 m_\pi^2+47 m_P^4 m_\pi^4-5 m_P^2 m_\pi^6\right)\nonumber\\
&+&q_y \left(4 k_x^2 m_P^2 \left(30 m_P^4-12 m_P^2 m_\pi^2+m_\pi^4\right)+6 m_P^6 m_\pi^2-2 m_P^4 m_\pi^4\right)+6 k_y^3 m_P^2 \left(10 m_P^4-6 m_P^2 m_\pi^2+m_\pi^4\right)\nonumber\\
&-& \left.8 k_y^2 m_P^2 q_y \left(30 m_P^4-12 m_P^2 m_\pi^2+m_\pi^4\right)\right)-4 k_x k_y^2 m_P^2 q_y \left(10 m_P^4-6 m_P^2 m_\pi^2+m_\pi^4\right)\nonumber\\
&+& q_x^2 \left(-8 k_x k_y^2 q_y \left(30 m_P^4-12 m_P^2 m_\pi^2+m_\pi^4\right)+k_x k_y \left(-240 m_P^6+96 m_P^4 m_\pi^2-8 m_P^2 m_\pi^4\right)\right.\nonumber\\
&-&\left. 4 k_x m_P^2 q_y \left(10 m_P^4-6 m_P^2 m_\pi^2+m_\pi^4\right)\right)+k_y \left(k_x \left(6 m_P^6 m_\pi^2-2 m_P^4 m_\pi^4\right)\right.\nonumber\\
&+&\left. k_x q_y^2 \left(120 m_P^6-48 m_P^4 m_\pi^2+4 m_P^2 m_\pi^4\right)\right)+k_x q_y \left(34 m_P^6 m_\pi^2-13 m_P^4 m_\pi^4+m_P^2 m_\pi^6\right)\nonumber\\
&+&\left. q_x^3 \left(4 k_y^3 \left(30 m_P^4-12 m_P^2 m_\pi^2+m_\pi^4\right)+k_y \left(60 m_P^6-36 m_P^4 m_\pi^2+6 m_P^2 m_\pi^4\right)\right)\right]\nonumber\\
&-&\left.12 m_\pi \left[\text{Li}_2\left(\frac{2 m_\pi^2}{m_\pi^2-\sqrt{m_\pi^4-4 m_P^2 m_\pi^2}}\right)+\text{Li}_2\left(\frac{2 m_\pi^2}{m_\pi^2+\sqrt{m_\pi^4-4 m_P^2 m_\pi^2}}\right)\right] (k_x q_y+k_y q_x)\right\},
\end{eqnarray}
and the remaining functions are
\begin{eqnarray}
\mathcal{K}_{k_x}& =&\mathcal{K}_{q_x}(k_x\leftrightarrow q_x, k_y\leftrightarrow q_y),
\\
\mathcal{K}_{k_y}& =&\mathcal{K}_{q_x}(k_x\leftrightarrow q_y, k_y\leftrightarrow q_x),
\\
\mathcal{K}_{q_y}& =&\mathcal{K}_{q_x}(k_x\leftrightarrow k_y, q_x\leftrightarrow q_y),
\\
\mathcal{K}_{q_y k_x}& =&\mathcal{K}_{q_x k_y}(k_x\leftrightarrow k_y, q_x\leftrightarrow q_y).
\end{eqnarray}
\end{subequations}

\subsubsection{Small-pion-mass approximation}
\label{subsec:Ks-SP}

By employing the small-pion-mass approximation, $m_\pi \ll m_P$, the expressions for $\mathcal{K}_{s}$ derived in the preceding subsection reduce to
\begin{subequations}
\label{K_smpi}
\begin{equation}
\mathcal{K}_0^{\text{SP}} \simeq \frac{-14m_P^4 - 8(k_y q_x - k_x q_y)^2 + 8(\mathbf{q}_\perp^2+\mathbf{k}_\perp^2+\mathbf{q}_\perp \cdot \mathbf{k}_\perp)m_P^2}{105m_P^{10}},
\end{equation}
\begin{equation}
\mathcal{K}_{q_x}^{\text{SP}} \simeq  \frac{-28 m_P^4 q_x - 8 q_x (k_y q_x - k_x q_y)^2 + 8q_x (\mathbf{q}_\perp^2+\mathbf{k}_\perp^2+\mathbf{q}_\perp \cdot \mathbf{k}_\perp)m_P^2+(2k_y+q_y)(k_y q_x -k_x q_y)m_P^2}{105m_P^{10}}, 
\end{equation}
\begin{eqnarray}
\mathcal{K}_{q_x k_x}^{\text{SP}}&\simeq& \frac{1}{315 m_P^{10}}\left\{-24 k_x^2 q_x \left[k_y \left(q_y^2-m_P^2\right)+m_P^2 q_y\right]-2 k_x \left[24 k_y^2 q_y (m_P^2-q_x^2)+3 k_y m_P^2 \left(7 m_P^2-8 q_x^2+4 q_y^2\right)-7 m_P^4 q_y\right.\right.\nonumber\\
&+&\left.\left. 24 m_P^2 q_x^2 q_y\right]+2 q_x \left[12 k_y^3 \left(3 m_P^2-q_x^2\right)+24 k_y^2 m_P^2 q_y-119 k_y m_P^4+12 k_y m_P^2 \left(3 q_x^2+q_y^2\right)-21 m_P^4 q_y\right]\right\}. 
\end{eqnarray}
\end{subequations}

\subsubsection{Center-of-mass frame}
\label{subsec:Ks-CM}

Upon imposing the center-of-mass condition, $\mathbf{k} = -\mathbf{q}$, the functions $\mathcal{K}_i$ simplify to
\begin{subequations}
\begin{eqnarray}
\mathcal{K}_0^{\text{CM}} &=& \frac{8}{3m_P^4 m_\pi^7} \Bigg\{2m_\pi^3 \mathbf{q}_\perp^2 +m_P^2(9m_\pi^3 - 60 m_\pi \mathbf{q}_\perp^2)-\frac{6m_P^2}{\sqrt{4m_P^2-m_\pi^2}}
\arctan\left(\frac{m_\pi}{\sqrt{4m_P^2-m_\pi^2}}\right) \nonumber\\
&& \times \left[m_P^2(6m_\pi^2-40\mathbf{q}_\perp^2)+8m_\pi^2 \mathbf{q}_\perp^2-m_\pi^4\right]\Bigg\},
\end{eqnarray}
\begin{eqnarray}
\mathcal{K}_{q_x}^{\text{CM}} &=& \frac{16 q_x}{3 m_P^4 m_\pi^7} \Bigg\{m_P^2 \left[9 m_\pi^3-30 m_\pi \mathbf{q}_\perp^2\right]+m_\pi^3 \mathbf{q}_\perp^2-\frac{6 m_P^2 }{\sqrt{4 m_P^2-m_\pi^2}} \arctan\left(\frac{m_\pi}{\sqrt{4 m_P^2-m_\pi^2}}\right) \nonumber\\
&&\times \left[m_P^2 \left(6 m_\pi^2-20 \mathbf{q}_\perp^2\right)-m_\pi^4+4 m_\pi^2 \mathbf{q}_\perp^2\right]\Bigg\} ,
\end{eqnarray}
\begin{eqnarray}
\mathcal{K}_{q_x k_y}^{\text{CM}} &=&\frac{8 q_x q_y}{3 m_P^4 m_\pi^7 \sqrt{4 m_P^2-m_\pi^2}} \left\{60 m_P^2 m_\pi \mathbf{q}_\perp^2 \sqrt{4 m_P^2-m_\pi^2}+48 m_P^2 m_\pi^2 \mathbf{q}_\perp^2 \arctan\left(\frac{m_\pi}{\sqrt{4 m_P^2-m_\pi^2}}\right)\right.\nonumber\\
&-&24 m_P^2 m_\pi^4 \arctan\left(\frac{m_\pi}{\sqrt{4 m_P^2-m_\pi^2}}\right)-2 m_\pi^3 \mathbf{q}_\perp^2 \sqrt{4 m_P^2-m_\pi^2}-240 m_P^4 \mathbf{q}_\perp^2 \arctan\left(\frac{m_\pi}{\sqrt{4 m_P^2-m_\pi^2}}\right)\nonumber\\
&-&39 m_P^2 m_\pi^3 \sqrt{4 m_P^2-m_\pi^2}+132 m_P^4 m_\pi^2 \arctan\left(\frac{m_\pi}{\sqrt{4 m_P^2-m_\pi^2}}\right)\nonumber\\
&+&\left.12 m_P^4 m_\pi \sqrt{4 m_P^2-m_\pi^2} \left[\text{Li}_2\left(\frac{2 m_\pi^2}{m_\pi^2-\sqrt{m_\pi^4-4 m_P^2 m_\pi^2}}\right)+ \text{Li}_2\left(\frac{2 m_\pi^2}{m_\pi^2+\sqrt{m_\pi^4-4 m_P^2 m_\pi^2}}\right)\right]\right\},
\end{eqnarray}
and the remaining functions are
\begin{eqnarray}
\mathcal{K}_{k_x}^{\text{CM}} &=&  - \mathcal{K}_{q_x}^{\text{CM}},
\\
\mathcal{K}_{k_y}^{\text{CM}} &=&  - \mathcal{K}_{q_y}^{\text{CM}},
\\
\mathcal{K}_{q_y}^{\text{CM}}&=&\mathcal{K}_{q_x}^{\text{CM}}(q_x\leftrightarrow q_y),
\\
\mathcal{K}_{q_x k_y}^{\text{CM}} &=&\mathcal{K}_{q_y k_x}^{\text{CM}}.
\end{eqnarray}
\end{subequations}

\subsubsection{Small-pion-mass approximation in center-of-mass frame}
\label{subsec:Ks-SC}

By applying the small-pion-mass approximation to the results obtained in the center-of-mass frame in Sec.~\ref{subsec:Ks-CM}, we derive
\begin{subequations}
\begin{eqnarray}
\mathcal{K}_0^{\text{CS}} &\simeq& \frac{8 \mathbf{q}_\perp^2 -14 m_P^2}{105 m_P^{8}} ,
\\
\mathcal{K}_{q_x}^{\text{CS}} &=&-\mathcal{K}_{k_x}^{\text{CS}} \simeq\frac{4q_x(2\mathbf{q}_\perp^2-7m_P^2)}{105 m_P^{8}},
\\
 \mathcal{K}_{q_y}^{\text{CS}}&=&-\mathcal{K}_{k_y}^{\text{CS}}= \mathcal{K}_{q_x}^{\text{CS}}(q_x\leftrightarrow q_y),
\\
 \mathcal{K}_{q_x k_y}^{\text{CS}}&=& \mathcal{K}_{q_y k_x}^{\text{CS}}\simeq \frac{4q_x q_y(35m_P^2-6\mathbf{q}_\perp^2)}{315m_P^8}.
\end{eqnarray}
\end{subequations}
Then we get
\begin{eqnarray}
\label{Final-F02}
\mathcal{F}_{0,2}^{\text{All},\text{CS}}&\simeq&  -\frac{m_P  |e B|^2}{2(2\pi)^2} \epsilon^{\mu\nu\alpha\beta}\left[k_{\parallel,\alpha} q_{\parallel,\beta}\frac{8 \mathbf{q}_\perp^2 -14 m_P^2}{105 m_P^{8}} +k_{\parallel,\alpha} q_{\perp,\beta}\frac{4(2\mathbf{q}_\perp^2-7m_P^2)}{105 m_P^{8}}\right.\nonumber\\
&+&\left.q_{\parallel,\beta} k_{\perp,\alpha}\frac{4(2\mathbf{q}_\perp^2-7m_P^2)}{105 m_P^{8}}
+k_{\perp,\alpha}q_{\perp,\beta}\frac{4(35m_P^2-6\mathbf{q}_\perp^2)}{315m_P^8}\right].
\end{eqnarray}

\section{Definition and calculations of $\mathcal{J}_i$ functions}
\label{J-Cal}

To evaluate the functions $\mathcal{J}_i$ (with $i=1,2,3$) appearing in Eq.~(\ref{F20A-F}), we employ the standard Feynman parametrization,
\begin{equation}
\frac{1}{A_1^{m_1}A_2^{m_2}\cdots A_n^{m_n}}=\int_0^1 dx_1\cdots dx_m\delta(\sum x_i-1)\frac{\prod x_i^{m_i-1}}{[\sum x_i A_i]^{\sum m_i}}\frac{\Gamma(m_1+\cdots +m_n)}{\Gamma(m_1)\cdots \Gamma(m_n)}.
\end{equation}
In particular, this yields the following relations:
\begin{eqnarray}
 \frac{1}{(s^2-m_P^2)[(s+q)^2-m_P^2][(s-k)^2-m_P^2]}&=&\int_{0}^{1} dx dy dz \delta(x+y+z-1) \frac{2}{D^3},
\\
  \frac{1}{(s^2-m_P^2)^4[(s+q)^2-m_P^2][(s-k)^2-m_P^2]} &=&\int_{0}^{1} dx dy dz \delta(x+y+z-1) \frac{20 x^3}{D^6},
\end{eqnarray}
where we have introduced the shorthand notation
\begin{equation}
 D = [x(s^2-m_P^2)+y((s+q)^2-m_P^2)+z((s-k)^2-m_P^2)]+i\epsilon = s^2+2s\cdot(yq -z k) + y q^2 + z k^2 -m_P^2 +i \epsilon.
\end{equation}
By defining a shifted integration variable $l = s + y q - z k$, the denominator can be written as $D = l^2 - \Delta + i\epsilon$, with $\Delta = (yq - zk)^2 - y q^2 - z k^2 +m_P^2$. Applying the on-shell conditions, $q^2 = k^2 = 0$ and $2q\cdot k = m_\pi^2$, the expression for $\Delta$ simplifies to $\Delta =- y z m_\pi^2 + m_P^2$.
Next, performing a Wick rotation, $l^0 \equiv i l_E^0$, we derive the following generic result:
\begin{equation}
\int \frac{d^4l}{(2\pi)^4}\frac{1}{\left(l^2-\Delta\right)^m} =i \int \frac{d^4 l_E}{(2\pi)^4} \frac{(-1)^m}{\left(l_E^2+\Delta\right)^m},
\end{equation}
where $D_E = l_E^2 + \Delta $.

By definition, the functions $\mathcal{J}_i$ in Eq.~(\ref{F20A-F}) are given by the following integral expressions:
\begin{subequations}
\begin{eqnarray}
\mathcal{J}_1&=&\int\frac{d^4 s}{(2\pi)^4}  \frac{\mathbf{s}_{\perp}^2}{(s^2-m_P^2)^4[(s+q)^2-m_P^2][(s-k)^2-m_P^2]}\nonumber\\
 &=&i\int \frac{dl_E^4}{(2\pi)^4}\int_0^1 dxdydz \delta(x+y+z-1)\frac{20x^3 [\mathbf{l}_\perp^2 + (z \mathbf{k}_\perp - y q_{\perp})^2] }{D_E^6},
\\
\mathcal{J}_2&=&\int\frac{d^4 s}{(2\pi)^4}  \frac{(\mathbf{s}_{\perp}+\mathbf{q}_\perp)^2}{(s^2-m_P^2)[(s+q)^2-m_P^2]^4[(s-k)^2-m_P^2]}\nonumber\\
 &=&i\int \frac{dl_E^4}{(2\pi)^4}\int_0^1 dxdydz \delta(x+y+z-1)\frac{20y^3 \{\mathbf{l}_\perp^2 + [z\mathbf{k}_\perp+(1-y)\mathbf{q}_\perp]^2\}}{D_E^6},
\\
\mathcal{J}_3&=&\int\frac{d^4 s}{(2\pi)^4}  \frac{(\mathbf{s}_{\perp}-\mathbf{k}_{\perp})^2}{(s^2-m_P^2)[(s+q)^2-m_P^2][(s-k)^2-m_P^2]^4}\nonumber\\
 &=&i\int \frac{dl_E^4}{(2\pi)^4}\int_0^1 dxdydz \delta(x+y+z-1)\frac{20z^3 \{\mathbf{l}_\perp^2 + [(1-z)\mathbf{k}_\perp + y \mathbf{q}_{\perp}]^2\} }{D_E^6},
\end{eqnarray}
\end{subequations}
where we have taken into account that $\mathbf{s}_\perp = \mathbf{l}_\perp- y \mathbf{q}_\perp + z \mathbf{k}_\perp$, $\mathbf{s}_\perp + \mathbf{q}_\perp = \mathbf{l}_\perp +(1- y)  \mathbf{q}_\perp + z \mathbf{k}_\perp$, $\mathbf{s}_\perp-\mathbf{k}_\perp =\mathbf{ l}_\perp - y \mathbf{q}_\perp -(1- z) \mathbf{k}_\perp$.

Each $\mathcal{J}_i$ can be conveniently separated into two parts:
\begin{subequations}
\begin{eqnarray}
\mathcal{J}_1^{1}&=&i\int \frac{dl_E^4}{(2\pi)^4}\int_0^1 dxdydz \delta(x+y+z-1)\frac{20x^3 \mathbf{l}_\perp^2}{D_E^6}
=i\int\frac{d l _E \pi^2}{(2\pi)^4} \int_{0}^{1} dx dy dz  \delta(x+y+z-1) \frac{20 x^3 l_E^5 }{D_E^6},
\\
\mathcal{J}_1^{2}&=&i\int \frac{dl_E^4}{(2\pi)^4}\int_0^1 dxdydz \delta(x+y+z-1)\frac{20x^3 (z \mathbf{k}_\perp - y q_{\perp})^2}{D_E^6}\nonumber\\
&=&i\frac{2\pi^2  }{3(2\pi)^4} \int_{0}^{1} dx dy dz  \delta(x+y+z-1) \frac{x^3(z \mathbf{k}_\perp - y \mathbf{q}_{\perp})^2 }{(-y z m_\pi^2+m_P^2)^4}, 
\end{eqnarray}
\begin{eqnarray}
\mathcal{J}_2^{1}&=&i\int \frac{dl_E^4}{(2\pi)^4}\int_0^1 dxdydz \delta(x+y+z-1)\frac{20y^3 \mathbf{l}_\perp^2}{D_E^6}
=i\int\frac{d l_E \pi^2}{(2\pi)^4} \int_{0}^{1} dx dy dz  \delta(x+y+z-1) \frac{20 y^3 l_E^5 }{D_E^6},
\\
\mathcal{J}_2^{2}&=&i\int \frac{dl_E^4}{(2\pi)^4}\int_0^1 dxdydz \delta(x+y+z-1)\frac{20y^3 \mathbf{q}_\perp^2 [z\mathbf{k}_\perp+(1-y)\mathbf{q}_\perp]^2}{D_E^6}\nonumber\\
&=&i\frac{2\pi^2  }{3(2\pi)^4} \int_{0}^{1} dx dy dz  \delta(x+y+z-1) \frac{y^3 [z\mathbf{k}_\perp+(1-y)\mathbf{q}_\perp]^2 }{(-y z m_\pi^2+m_P^2)^4}, 
\end{eqnarray}
\begin{eqnarray}
\mathcal{J}_3^{1}&=&i\int \frac{dl_E^4}{(2\pi)^4}\int_0^1 dxdydz \delta(x+y+z-1)\frac{20z^3 \mathbf{l}_\perp^2}{D_E^6}
=i\int\frac{d l_E \pi^2}{(2\pi)^4} \int_{0}^{1} dx dy dz  \delta(x+y+z-1) \frac{20 z^3 l_E^5 }{D_E^6},
\\
\mathcal{J}_3^{2}&=&i\int \frac{dl_E^4}{(2\pi)^4}\int_0^1 dxdydz \delta(x+y+z-1)\frac{20z^3 [(1-z)\mathbf{k}_\perp + y \mathbf{q}_{\perp}]^2}{D_E^6}\nonumber\\
&=&i\frac{2\pi^2 }{3(2\pi)^4} \int_{0}^{1} dx dy dz  \delta(x+y+z-1) \frac{z^3[(1-z)\mathbf{k}_\perp + y \mathbf{q}_{\perp}]^2 }{(-y z m_\pi^2+m_P^2)^4}. 
\end{eqnarray}
\end{subequations}
Next, performing the integrations and adding all separate contributions together, we obtain
\begin{equation}
\label{I1_F}
 \sum_{i=1}^{3}\mathcal{J}_i^{1}= \frac{i(6m_P^2-m_\pi^2)\left[m_\pi \sqrt{4m_P^2-m_\pi^2}(m_\pi^2-3m_P^2)+(12m_P^4-6m_P^2 m_\pi^2+m_\pi^4)\arctan\left(\frac{m_\pi}{\sqrt{4m_P^2-m_\pi^2}}\right)\right]}{48\pi^2 m_P^4 m_\pi^5(4m_P^2-m_\pi^2)^{\frac{3}{2}}}, 
 \end{equation}
 and
\begin{eqnarray}
\label{I2_F}
\sum_{i = 1}^{3}\mathcal{J}^2_i  &=&i\frac{ 1 }{216(2\pi)^2 m_P^6 m_\pi^7}\left\{\frac{1}{(4m_P^2-m_\pi^2)^2}\left[720 m_P^8 m_\pi \left(\mathbf{k}_\perp^2-6 \mathbf{k}_\perp \cdot \mathbf{q}_\perp+\mathbf{q}_\perp^2\right)-480 m_P^6 m_\pi^3 \left(\mathbf{k}_\perp^2-6 \mathbf{k}_\perp \cdot \mathbf{q}_\perp+\mathbf{q}_\perp^2\right)\right.\right.\nonumber\\
&+&\left. 4 m_P^4 m_\pi^5 \left(23 \mathbf{k}_\perp^2-156 \mathbf{k}_\perp \cdot \mathbf{q}_\perp +23 \mathbf{q}_\perp^2\right)+2 m_P^2 m_\pi^7 \left(\mathbf{k}_\perp^2+24 \mathbf{k}_\perp \cdot \mathbf{q}_\perp+\mathbf{q}_\perp^2\right)-m_\pi^9 \left(\mathbf{k}_\perp^2+\mathbf{q}_\perp^2\right)\right]\nonumber\\
&+&\frac{24 m_P^2}{(4m_P^2-m_\pi^2)^{\frac{5}{2}}} \left[-120 m_P^8 \left(\mathbf{k}_\perp^2-6 \mathbf{k}_\perp \cdot \mathbf{q}_\perp+\mathbf{q}_\perp^2\right)+100 m_P^6 m_\pi^2 \left(\mathbf{k}_\perp^2-6 \mathbf{k}_\perp \cdot \mathbf{q}_\perp+\mathbf{q}_\perp^2\right)\right.\nonumber\\
&-&\left. \left. 4 m_P^4 m_\pi^4 \left(7 \mathbf{k}_\perp^2-45 \mathbf{k}_\perp \cdot \mathbf{q}_\perp+7 \mathbf{q}_\perp^2\right)+m_P^2 m_\pi^6 \left(3 \mathbf{k}_\perp^2-22 \mathbf{k}_\perp \cdot \mathbf{q}_\perp +3 \mathbf{q}_\perp^2\right)+ m_\pi^8 \mathbf{k}_\perp \cdot \mathbf{q}_\perp \right]\arctan\left(\frac{m_\pi}{\sqrt{4m_P^2-m_\pi^2}}\right)\right\}.\nonumber\\
\end{eqnarray}

\subsubsection{$\mathcal{J}_i$ functions in the small-pion-mass approximation}

In the small-pion-mass approximation, the general results for $\mathcal{J}_i$ functions given Eqs.~(\ref{I1_F}) and (\ref{I2_F}) drastically simplify. 
The corresponding results read
\begin{subequations}	
\label{Jsp}
\begin{eqnarray}
\sum_{i = 1}^{3}\mathcal{J}_i^{1, \text{SP}} &\simeq& \frac{i}{320\pi^2 m_P^6},
\\
\sum_{i = 1}^{3}\mathcal{J}_i^{2, \text{SP}} &\simeq& i \frac{\mathbf{k}_\perp^2+ \mathbf{k}_\perp \cdot \mathbf{q}_\perp + \mathbf{q}_\perp^2}{2016\pi^2 m_P^8}.
\end{eqnarray}
\end{subequations}

\subsubsection{$\mathcal{J}_i$ functions in the center-of-mass frame}

In the center-of-mass frame, Eqs.~(\ref{I1_F}) and (\ref{I2_F}) reduce to the following expressions:
\begin{subequations}
\label{JCM}
 \begin{equation}
 \sum_{i=1}^{3}\mathcal{J}_i^{1,\text{CM}} = \frac{i(6m_P^2-m_\pi^2)\left[m_\pi \sqrt{4m_P^2-m_\pi^2}(m_\pi^2-3m_P^2)+(12m_P^4-6m_P^2 m_\pi^2+m_\pi^4)\arctan\left(\frac{m_\pi}{\sqrt{4m_P^2-m_\pi^2}}\right)\right]}{48\pi^2 m_P^4 m_\pi^5(4m_P^2-m_\pi^2)^{\frac{3}{2}}},
\end{equation}
\begin{equation}
\sum_{i = 1}^{3}\mathcal{J}^{2,\text{CM}}_i = i\frac{ \mathbf{q}_\perp^2 }{6(2\pi)^2}\left[\frac{180m_P^4 - 30 m_P^2 m_\pi^2 - m_\pi^4}{18 m_P^6 m_\pi^6}-\frac{2(60m_P^4-20m_P^2 m_\pi^2 + m_\pi^4)}{3m_P^4 m_\pi^7\sqrt{4m_P^4-m_\pi^2}}\arctan\left(\frac{m_\pi}{\sqrt{4m_P^2-m_\pi^2}}\right)\right].
\end{equation}
\end{subequations}

\subsubsection{$\mathcal{J}_i$ functions in the small-pion-mass approximation and center-of-mass frame}

From the above derivations, in the center-of-mass frame with a small-pion-mass limit, we have 
\begin{subequations}
\label{JCS}
\begin{eqnarray}
\sum_{i=1}^{3}\mathcal{J}_i^{1, \text{CS}}&=&\sum_{i=1}^{3}\mathcal{J}_i^{1, \text{SP}},  
\\
\sum_{i=1}^{3}\mathcal{J}_i^{2, \text{CS}} &\simeq& \frac{i \mathbf{q}_\perp^2}{2016\pi^2 m_P^8}. 
\end{eqnarray}
\end{subequations}

\end{document}